\documentclass[aps,prd,10pt,notitlepage,nofootinbib]{revtex4-1}

\usepackage[utf8]{inputenc}
\usepackage{amsmath,amssymb,amsfonts}
\usepackage{newtxtext,newtxmath}

\usepackage{bbold}
\usepackage{bm}
\usepackage{graphicx}
\usepackage[usenames,dvipsnames]{xcolor}
\usepackage{color}
\usepackage[colorlinks=true,linkcolor=blue,urlcolor=blue,citecolor=blue]{hyperref}
\usepackage{slashed}
\usepackage[english]{babel}
\usepackage{dcolumn}
\usepackage{pifont}
\usepackage{dsfont,mathrsfs}
\usepackage{cancel}
\usepackage{bigints}
\usepackage{accents}
\usepackage{soul}
\usepackage{multirow}

\usepackage{amsmath, amssymb}
\usepackage{bbold}
\usepackage{makecell}
\usepackage{simpler-wick}

\usepackage{orcidlink} %Added by Snigdha

\newcommand{\nn}{\nonumber}

\newcommand{\ensembleaverage}[1]{\left\langle#1\right\rangle}

\newcommand{\FB}[1]{\left(#1\right)}
\newcommand{\fb}[1]{(#1)}
\newcommand{\SB}[1]{\left\{#1\right\}}
\newcommand{\TB}[1]{\left[#1\right]}

\newcommand{\munu}{{\mu\nu}}

\newcommand{\IM}{\text{Im}}
\newcommand{\RE}{\text{Re}}
\newcommand{\Tr}{\text{Tr}}

\newcommand{\psibar}{\overline{\psi}}

\newcommand{\del}{\partial}
\newcommand{\identity}{\mathds{1}}

\newcommand{\wpr}{\omega^r_{\bm{p}}}
\newcommand{\wps}{\omega^s_{\bm{p}}}
\newcommand{\wkr}{\omega^r_{\bm{k}}}
\newcommand{\wks}{\omega^s_{\bm{k}}}
\newcommand{\fpkr}{f_+^{\bm{k}r}}
\newcommand{\fmkr}{f_-^{\bm{k}r}}
\newcommand{\fpps}{f_+^{\bm{p}s}}
\newcommand{\fmps}{f_-^{\bm{p}s}}
\newcommand{\fpks}{f_+^{\bm{k}s}}
\newcommand{\fmks}{f_-^{\bm{k}s}}

\allowdisplaybreaks

\begin{document}
%\begin{linenumbers}
\title{The mass and spectral function of scalar and pseudoscalar mesons in a hot and chirally imbalanced medium using the two-flavor NJL model}

	\author{Snigdha Ghosh\orcidlink{0000-0002-2496-2007}$^{a}$}
	\email{snigdha.ghosh@bangla.gov.in}
%	\email{snigdha@ggdckharagpur2.ac.in}
	\email{snigdha.physics@gmail.com}
	\thanks{Corresponding Author}

	\author{Nilanjan Chaudhuri\orcidlink{0000-0002-7776-3503}$^{b,d}$}
	\email{sovon.nilanjan@gmail.com}
	\email{n.chaudhri@vecc.gov.in}

	\author{Sourav Sarkar\orcidlink{0000-0002-2952-3767}$^{b,d}$}
	\email{sourav@vecc.gov.in}
	
	\author{Pradip Roy$^{c,d}$}
	\email{pradipk.roy@saha.ac.in}
	
	\affiliation{$^a$Government General Degree College Kharagpur-II, Paschim Medinipur - 721149, West Bengal, India}	
	\affiliation{$^a$Variable Energy Cyclotron Centre, 1/AF Bidhannagar, Kolkata - 700064, India}
	\affiliation{$^c$Saha Institute of Nuclear Physics, 1/AF Bidhannagar, Kolkata - 700064, India}
	\affiliation{$^d$Homi Bhabha National Institute, Training School Complex, Anushaktinagar, Mumbai - 400085, India}

	%+++++++++++++++++++++++++++++++++++++++++++++++++++++++++++++++++++++++++++++++++++++++++++++++++++++++++++++++++++++
\begin{abstract}
We explore the properties of neutral mesons within the context of a chirally imbalanced medium, employing the two-flavor Nambu--Jona-Lasinio model. The temperature dependence of the constituent quark mass at finite values of the chiral chemical potential (CCP) demonstrates the well-established phenomena of chiral catalysis at lower temperatures and inverse chiral catalysis at higher temperatures. The polarization functions  in both the scalar ($\sigma$) and pseudo-scalar ($\pi^0$) channels have been evaluated using real time formalism of thermal field theory. These have been used to determine the masses and spectral functions of $ \sigma $ and $ \pi $ mesons. Detailed investigation of the analytic structure of the imaginary part of the polarization function for $\sigma$ and $\pi$ mesons results in the emergence of non-trivial Landau cut contributions due to the presence of chiral imbalance. The multiple solutions for the mass of the $\pi$ meson for specific values of CCP have been analysed on the basis of their residue at the pole. Furthermore, we have observed abrupt changes in the masses of both scalar and pseudo-scalar mesons at finite CCP values, particularly at higher temperatures. A decreasing trend in the Mott transition temperature is seen with the increase in CCP.
\end{abstract}

\maketitle
\section{Introduction}

Quantum chromodynamics (QCD) is a non-Abelian gauge theory characterized by a gauge group $ SU(3) $. It provides a framework for studying the strong interactions occurring between quarks and gluons, originating through the exchange of color charges. The study of the QCD vacuum structure under extreme conditions of temperature and/or baryon density stands as a central objective in the context of relativistic heavy ion collision (HIC) experiments at RHIC and LHC. It is well established that the numerous energy-degenerate vacuum configurations of QCD at zero and low temperatures can be identified through topologically non-trivial gauge configurations with a non-zero winding number~\cite{Shifman:1988zk}. These particular gluon configurations are referred to as instantons, and they have the capability to induce transitions between distinct vacua by crossing potential barriers with heights approximately on the order of the QCD scale $\Lambda_\text{QCD}$. This phenomenon is known as instanton tunneling~\cite{Belavin:1975fg,tHooft:1976rip,tHooft:1976snw}.
However, at high temperatures, particularly in the quark-gluon plasma (QGP) phase observed in HICs, an abundant production of another class of gluon configurations, known as sphalerons, is anticipated~\cite{Manton:1983nd,Klinkhamer:1984di}. It is conjectured that the prevalence of sphalerons can enhance the transition rate across energy-degenerate vacuum states by surmounting the energy barriers~\cite{Kuzmin:1985mm,Arnold:1987mh,Khlebnikov:1988sr,Arnold:1987zg}.
The gauge field configurations with topological non-triviality have the capability to alter the helicities of quarks during interactions, thus leading to the violation of parity ($P$) and charge-parity ($CP$) symmetries by generating an asymmetry between left and right-handed quarks, as a consequence of the axial anomaly of QCD~\cite{Adler:1969gk,Bell:1969ts}. The production of chirality imbalance can occur locally, as there is no direct evidence of the global violation of $P$ and $CP$ in QCD~\cite{Adler:1969gk,Bell:1969ts,McLerran:1990de,Moore:2010jd}. This locally induced chirality imbalance is characterized using a chiral chemical potential (CCP), which essentially quantifies the difference between the numbers of right and left-handed quarks.

In non-central heavy-ion collisions (HICs), magnetic fields of very high intensity, of the order of a few $m_\pi^2$, can arise~\cite{Kharzeev:2007jp,Skokov:2009qp}. When such strong magnetic fields coexist with chirality imbalance, they can result in the separation of positive and negative charges with respect to the reaction plane, giving rise to an electric current along the magnetic field direction. This phenomenon is known as the chiral magnetic effect (CME)~\cite{Fukushima:2008xe,Kharzeev:2007jp,Kharzeev:2009pj,Bali:2011qj}.
Extensive efforts have been dedicated to the detection of the CME in HIC experiments conducted at RHIC at Brookhaven. However, a recent analysis by the STAR Collaboration has not provided any evidence of the CME occurring in these collisions~\cite{STAR:2021mii}. Consequently, new experimental techniques for detecting the CME have been proposed~\cite{An:2021wof,Milton:2021wku,Kharzeev:2022hqz,STAR:2022zpv}.

Given the anticipation of locally generated chirality imbalance in QGP, in addition to the CME, considerable research endeavours have been devoted to understand various facets of this phenomenon.  These studies encompass investigations into  microscopic transport phenomena ~\cite{Vilenkin:1979ui,Vilenkin:1980fu,Fukushima:2008xe,Son:2009tf}, collective oscillations ~\cite{Akamatsu:2013pjd,Carignano:2018thu,Carignano:2021mrn}, fermion damping rate~\cite{Carignano:2019ivp} and collisional energy loss of fermions~\cite{Carignano:2021mrn}  within chirally imbalanced media.
 Furthermore, chirally asymmetric plasmas are anticipated to form in the gap regions of the magnetospheres of pulsars and black holes~\cite{Gorbar:2021tnw}, as well as in other stellar astrophysical scenarios~\cite{Charbonneau:2009ax,Akamatsu:2013pjd,Kaminski:2014jda,Yamamoto:2015gzz,Shovkovy:2021yyw}. It's worth noting that the CME has been observed in condensed matter systems, particularly in 3D Dirac and Weyl semimetals~\cite{Li:2014bha,Li:2016vlc,Kharzeev:2013ffa,Kharzeev:2015znc,Huang:2015oca,Landsteiner:2016led,Gorbar:2017lnp,Joyce:1997uy,Tashiro:2012mf}. As a result, the investigation of the properties of chirally imbalanced matter remains a topic of significant and ongoing interest.

However, owing to the non-perturbative nature of QCD it is difficult to address these problems from first principles calculations. Most of our current understanding comes from Lattice QCD (LQCD) simulations~\cite{deForcrand:2006pv,Aoki:2006br,Aoki:2009sc,Bazavov:2009zn,Cheng:2007jq,Muroya:2003qs}. But, direct simulation of QCD thermodynamics at finite density is not possible, due to the sign  problem. This arises when simulating QCD at non-zero baryon density because the fermion determinant, which is necessary for calculating observables, becomes complex and prevents the application of traditional Monte Carlo methods. An alternative for probing low-energy QCD involves the utilization of effective models. These models aim to isolate the relevant physics of a phenomenon by constructing mathematically manageable frameworks which can be used to probe the regions that are beyond the scope of perturbative QCD or LQCD. Often, these effective models are parametrized through fitting with more reliable theoretical frameworks or experimental data.
In this work we will employ the Nambu$ - $Jona-Lasinio (NJL) model~\cite{Nambu:1961fr,Nambu:1961tp} which  respects the global symmetries of QCD, most importantly the chiral symmetry. Being an effective model,  within its domain of applicability it provides a useful scheme to study some of the important non-perturbative properties of the QCD vacuum ~\cite{Klevansky:1992qe,Vogl:1991qt,Buballa:2003qv,Volkov:2005kw}. As the gluonic degrees of freedom have been integrated out in this effective description,  point-like interaction among the quarks  appears~\cite{Klevansky:1992qe} which makes the NJL  model non-renormalizable. Therefore, one has to choose a proper regularization scheme  to  deal with the divergent integrals and subsequently, fix the model parameters by  reproducing  a set of   phenomenological quantities, for example the pion-decay constant, quark condensate \textit{etc}. 
This model has been extensively used to study  the phase structure~\cite{Ruggieri:2016lrn,Ruggieri:2016asg,Ruggieri:2020qtq,Chaudhuri:2021lui} as well as properties of electromagnetic spectral function~\cite{Ghosh:2022xbf}, dilepton production rate~\cite{Chaudhuri:2022rwo} in a chirally asymmetric medium.

It is well known that, mesons are the bound/resonant states of quarks and antiquarks, so their propagation can be studied from the scattering of quarks/antiquarks in different channels employing the Bethe-Salpeter formalism. The physical properties of light scalar and pseudoscalar mesons have a direct relevance with the dynamics of chiral phase transition. In literature a lot of works can be found where NJL model is used to study different mesonic modes at finite temperature and/or density~\cite{Hatsuda:1985eb,Ayala:2002qy,Hansen:2006ee,Inagaki:2007dq}. However, we have not encountered any references that investigates spectral properties of mesons in the presence of chiral imbalance. Polarization functions of mesons play a crucial role in the computation of meson propagators within the framework of the Random Phase Approximation (RPA)~\cite{Klevansky:1992qe}. These propagators are instrumental in calculating quark and anti-quark in medium scattering cross-sections, which are essential for determining various transport coefficients using NJL-like models~\cite{Zhuang:1995uf} which involve meson exchange. Consequently, these polarization functions facilitate the study of dissipative phenomena in strongly interacting matter in presence of chiral imbalance.

The paper is organized as follows. In Sec.~\ref{sec.gap} we have discussed the gap equation in the presence of CCP. The meson propagators and spectral functions are derived in Sec.~\ref{sec.propagator}. We present numerical results and related discussions in Sec.~\ref{sec.results} followed by summary and conclusions in Sec.~\ref{sec.summary}.  

%~~~~~~~~~~~~~~~~~~~~~~~~~~~~~~~~~~~~~~~~~~~~~~~~~~~~~~~~~~~~~~~~~~~~~~~~~~~~~~~~~~~~~~~~~~~~~~~	
\section{Gap Equation at Finite CCP} \label{sec.gap}
Let us start with the standard expression of the Lagrangian (density) for a 2-flavor NJL model at finite CCP $\mu_5$ and quark chemical potential $\mu$ as~\cite{Nambu:1961fr,Nambu:1961tp,Ghosh:2022xbf,Chaudhuri:2022rwo}
\begin{eqnarray}
	\mathscr{L}_\text{NJL} = \psibar(i\cancel{\del}-m+\mu\gamma^0+\mu_5\gamma^0\gamma^5)\psi + G_s\TB{(\psibar\psi)^2+(\psibar i\gamma^5\bm{\tau}\psi)^2} \label{lagrangian}
\end{eqnarray} 
where $\psi = \FB{u~d}^T$ is the quark iso-doublet, $m$ is the current quark mass, $\bm{\tau}$ is the Pauli isospin matrices and $G_s$ represents the scalar coupling. Throughout the article, we will be using metric tensor $g^\munu=\text{diag}(1,-1,-1,-1)$.

In the Hartree mean field approximation (MFA), the constituent quark mass $M$ is obtained by solving the gap equation
\begin{eqnarray}
	M = m - 2G_s\ensembleaverage{\psibar \psi} \label{gap}
\end{eqnarray}
where, the chiral condensate is given by
\begin{eqnarray}
	\ensembleaverage{\psibar \psi} = \RE~i\int\frac{d^4p}{(2\pi)^4} \Tr_\text{d,c,f}\TB{S_{11}(p)} \label{condensate}
\end{eqnarray}
in which the $\Tr_\text{d,c,f}[...]$ implies taking trace in Dirac, color and flavor spaces and  $S_{11}(p)$ is the $11$ component of the real time thermal quark propagator defined in the context of real time formulation (RTF) of finite temperature field theory. The explicit expression of $S_{11}(p)$ is given by~\cite{Ghosh:2022xbf,Chaudhuri:2022rwo}
\begin{eqnarray}
	S_{11}(p) = \mathcal{D}(p^0+\mu,\bm{p}) \sum_{r\in\{\pm\}}^{}\frac{1}{4|\bm{p}|r\mu_5}\TB{\frac{-1}{(p^0+\mu)^2-(\wpr)^2+i\epsilon}-2\pi i\eta(p^0+\mu)\delta\SB{(p^0+\mu)^2-(\wpr)^2}}\otimes\identity_\text{color}\otimes\identity_\text{flavor}
	\label{S11}
\end{eqnarray}
where, $r$ represents the helicity of the propagating quark, $\wpr = \sqrt{(|\bm{p}|+r\mu_5)^2+M^2}$ is the single particle dispersion in presence of CCP, $\eta(x)=\theta(x)f_+(x)+\theta(-x)f_-(-x)$, $\theta(x)$ is the unit step function and $f_\pm(x) = \dfrac{1}{e^{(x\mp\mu)/T}+1}$ is the Fermi-Dirac thermal distribution function and $\mathcal{D}$ contains complicated Dirac structure as follows
\begin{eqnarray}
	\mathcal{D}(p^0,\bm{p}) = \sum_{j \in \{\pm\}} \mathscr{P}_j \TB{ p_{-j}^2\cancel{p}_j - M^2 \cancel{p}_{-j} + M (p_j\cdot p_{-j}-M^2) + i M \sigma_\munu p_j^\mu p_{-j}^\nu  } \label{D}
\end{eqnarray}
in which $\mathscr{P}_j = \frac{1}{2}(\identity+j\gamma^5)$ and $p^\mu_j \equiv (p^0 +j\mu_5,\bm{p})$. Substituting the propagator from Eq.~\eqref{S11} into Eq.~\eqref{condensate} followed by performing the $dp^0$ integration, we get after  some simplifications the chiral condensate as
\begin{eqnarray}
	\ensembleaverage{\psibar\psi} = -N_c N_f M \sum_{r \in \{\pm\}} \int \frac{d^3p}{(2\pi)^3} \frac{1}{\wpr} \FB{1-f_+^{\bm{p}r} -f_-^{\bm{p}r}}
	\label{condensate.2}
\end{eqnarray}
where, $N_c=3$ is the number of colors, $N_f=2$ is the number of flavors and $f_\pm^{\bm{p}r} = f_\pm(\wpr)$. The first term within the parenthesis in the above integral is the temperature independent vacuum contribution to the chiral condensate which is UV divergent; and as the NJL model is known to be non-renormalizable, a proper regularization prescription has to be employed to tackle the divergences. Various regularization schemes have been explored in the literature. One of the widely adopted approaches involves the use of a physically intuitive three-momentum cutoff, which is noncovariant in nature. This cutoff can be applied in two ways: by employing the regulator solely in the temperature-independent integrals or by applying it in both the vacuum and thermal parts, even if the thermal part is already UV finite. Additionally, several covariant schemes have been introduced, such as the four-momentum cutoff in Euclidean space, regularization in proper time, and the Pauli-Villars method. A comprehensive discussion on different regularization schemes can be found in references such as~\cite{Klevansky:1992qe, Kohyama:2015hix, Xue:2021ldz}. In this work we will be using the smooth three-momentum cutoff regulator only in the divergent vacuum term following Ref.~\cite{Fukushima:2010fe} as the sharp momentum cutoff suffers from a regularization artifact. This is implemented by replacing the vacuum term as
\begin{eqnarray}
	\int\dfrac{d^3p}{(2\pi)^3}1 \to \int\dfrac{d^3p}{(2\pi)^3} f_\Lambda^{\bm{p}} \label{regularization}
\end{eqnarray}
where $f_\Lambda^{\bm{p}} = \sqrt{\dfrac{\Lambda^{20}}{\Lambda^{20} + |\bm{p}|^{20}}}$. 
Substituting Eq.~\eqref{condensate.2} in Eq.~\eqref{gap} and making use of Eq.~\eqref{regularization}, we finally obtain the gap equation as
\begin{eqnarray}
	M = m + 2G_s N_c N_f M \sum_{r \in \{\pm\}} \int \frac{d^3p}{(2\pi)^3} \frac{1}{\wpr} \FB{f_\Lambda^{\bm{p}} -f_+^{\bm{p}r} -f_-^{\bm{p}r}}. \label{gap.final}
\end{eqnarray}

%~~~~~~~~~~~~~~~~~~~~~~~~~~~~~~~~~~~~~~~~~~~~~~~~~~~~~~~~~~~~~~~~~~~~~~~~~~~~~~~~~~~~~~~~~~~~~~~~
%
\section{MESON PROPAGATORS AND SPECTRAL FUNCTIONS} \label{sec.propagator}
Mesons are the bound/resonant states of quarks and antiquarks, so their propagation can be studied from the scattering of quarks/antiquarks in different channels using the Bethe-Salpeter formalism~\cite{Hatsuda:1985eb,Klevansky:1992qe,Ayala:2002qy,Hansen:2006ee,Inagaki:2007dq,Ghosh:2020qvg}. Using the random phase approximation (RPA), the meson propagators $D_{h}$ in the scalar ($\pi$) and pseudo-scalar ($\sigma$) channels can be written as 
\begin{eqnarray}
	D_h(q;T,\mu_5) = \frac{2G}{1-2G\Pi_h(q;T,\mu_5)} ~~;~~ h\in \{\pi,\sigma\} \label{Dh}
\end{eqnarray}
where, $\Pi_h$ is the one-loop polarization function of meson $h$.

The one-loop  polarization function $\Pi_h(q)$ at finite CCP and temperature can be calculated using the RTF of finite temperature field theory. Being a two-point correlation function, the real time polarization function acquires a $2\times2$ matrix structure (in thermal space) whose $11$-component is given by
\begin{eqnarray}
	\Pi_h^{11}(q) = i\int\!\!\frac{d^4k}{(2\pi)^4}\Tr_\text{d,c,f}\TB{\Gamma_hS_{11}(p=q+k)\Gamma_hS_{11}(k)} \label{Pi.11}
\end{eqnarray}
where, $h\in \{\pi,\sigma\}$, $\Gamma_\pi = i\gamma^5$, $\Gamma_\sigma = \identity$ and the quark propagator $S_{11}(p)$ is defined in Eq.~\eqref{S11}. Now substituting Eq.~\eqref{S11} into Eq.~\eqref{Pi.11} and by performing the $dk^0$ integration, we obtain the real and imaginary parts of the 11-component of mesonic polarization function as
\begin{eqnarray}
	\RE\Pi_h^{11}(q_0,\bm{q}) &=& N_cN_f \int \frac{d^3k}{(2\pi)^3} \sum_{r \in \{\pm\}} \sum_{s \in \{\pm\}} 
	\frac{1}{64rs\mu_5^2 |\bm{p}||\bm{k}|} 
	\mathcal{P} \Big[ \frac{\mathcal{N}_h(k^0=\wkr)(f_\Lambda^{\bm{k}}-2\fpkr)}{\wkr \SB{(q^0+\wkr)^2-(\wps)^2}}	
	+ \frac{\mathcal{N}_h(k^0=-\wkr)(f_\Lambda^{\bm{k}}-2\fmkr)}{\wkr \SB{(q^0-\wkr)^2-(\wps)^2}} \nn \\	
	&& + \frac{\mathcal{N}_h(k^0=-q^0+\wps)(f_\Lambda^{\bm{k}}-2\fpps)}{\wps \SB{(q^0-\wps)^2-(\wkr)^2}}
	+ \frac{\mathcal{N}_h(k^0=-q^0-\wps)(f_\Lambda^{\bm{k}}-2\fmps)}{\wps \SB{(q^0+\wps)^2-(\wkr)^2}} \Big], \label{RePi.1} \\
	\IM\Pi_h^{11}(q_0,\bm{q}) &=& -N_cN_f \pi \int \frac{d^3k}{(2\pi)^3} \sum_{r \in \{\pm\}} \sum_{s \in \{\pm\}} 
	\frac{1}{16rs\mu_5^2 |\bm{p}||\bm{k}|} \frac{1}{4\wkr\wps} \nn \\ && \hspace{-2.5cm}
	\times \Big[ \mathcal{N}_h(k^0=-\wkr) \fb{1-\fmkr -\fpps +2\fmkr\fpps}\delta(q_0-\wkr-\wps) 
	+ \mathcal{N}_h(k^0=\wkr)  \fb{1-\fpkr -\fmps +2\fpkr\fmps} \delta(q_0+\wkr+\wps) \nn \\ && \hspace{-2.5cm}
	+ \mathcal{N}_h(k^0=\wkr) \fb{-\fpkr -\fpps +2\fpkr\fpps} \delta(q_0+\wkr-\wps) 
	+ \mathcal{N}_h(k^0=-\wkr) \fb{-\fmkr -\fmps +2\fmkr\fmps} \delta(q_0-\wkr+\wps) \Big]
	\label{ImPi.1}
\end{eqnarray}
where, the smooth-cutoff $f_\Lambda^{\bm{k}}$ has been used to regulate the UV-divergent part of $\RE\Pi_h^{11}(q_0,\bm{q})$, $\mathcal{P}$ denotes Cauchy principal value integration, and
\begin{eqnarray}
	\mathcal{N}_h(k) &=& 4a_hM^6 + 2M^4 \big\{-2a_h (k_+\cdot k_-)+(k_+\cdot p_-)+(k_-\cdot p_+)-2a_h(p_+\cdot p_-)\big\}-2M^2 \big\{ -2a_h(k_+\cdot p_-)(k_+\cdot p_+) \nn \\ 
	&& -2a_h(k_+\cdot k_-)(p_+\cdot p_-) + 2a_h(k_+\cdot p_+)(k_-\cdot p_-) + (k_-\cdot p_-)(k_+^2+p_+^2) + (k_+\cdot p_+)(k_-^2+p_-^2) \big\} \nn \\
	&& + 2 \big\{ (k_+\cdot p_-)k_-^2p_+^2 + (k_-\cdot p_+)k_+^2p_-^2 \big\} \label{Nh}
\end{eqnarray} 
in which $a_\pi=-1$ and $a_\sigma=1$. Having calculated the 11-components, it is now easy to diagonalize the real-time thermal polarization matrix~\cite{Mallik:2016anp}, and extract the diagonal components which are analytic functions that appear in Eq.~\eqref{Dh}. The real and imaginary parts of the analytic function $\Pi_h(q)$ are related to the 11-components via relations
\begin{eqnarray}
	\RE\Pi_h(q) = \RE\Pi_h^{11}(q) ~~\text{and}~~~ \IM\Pi_h(q) = \text{sign}(q^0)\tanh\FB{\frac{q^0}{2T}}\IM\Pi_h^{11}(q) \label{bar.to.11}
\end{eqnarray}
respectively. For the calculation of meson masses, we only require the special cases: $\Pi_h(q^0,\bm{q}=\bm{0})$. For $q^0\ne0,\bm{q}=\bm{0}$, substitution of Eqs.~\eqref{RePi.1} and \eqref{ImPi.1} into Eq.~\eqref{bar.to.11} yields after  simplification: 
\begin{eqnarray}
	\RE\Pi_h(q_0,\bm{q}=\bm{0}) &=& \frac{N_cN_f}{2\pi^2} \int_{0}^{\infty}d|\bm{k}| \sum_{r \in \{\pm\}} \sum_{s \in \{\pm\}} 
	\frac{1}{64rs\mu_5^2} 
	\mathcal{P}\Big[ \frac{\mathcal{N}_h(k^0=\wkr)(f_\Lambda^{\bm{k}}-2\fpkr)}{\wkr \SB{(q^0+\wkr)^2-(\wks)^2}}	
	+ \frac{\mathcal{N}_h(k^0=-\wkr)(f_\Lambda^{\bm{k}}-2\fmkr)}{\wkr \SB{(q^0-\wkr)^2-(\wks)^2}} \nn \\	
	&& + \frac{\mathcal{N}_h(k^0=-q^0+\wks)(f_\Lambda^{\bm{k}}-2\fpks)}{\wks \SB{(q^0-\wks)^2-(\wkr)^2}}
	+ \frac{\mathcal{N}_h(k^0=-q^0-\wks)(f_\Lambda^{\bm{k}}-2\fmks)}{\wks \SB{(q^0+\wks)^2-(\wkr)^2}} \Big]_{\bm{q}=\bm{0}}, 
	\label{RePi.2} \\
	\IM\Pi_h(q_0,\bm{q}=\bm{0}) &=& - \frac{N_cN_f}{2\pi} \int_{0}^{\infty}d|\bm{k}| \sum_{r \in \{\pm\}} \sum_{s \in \{\pm\}} 
	\frac{1}{16rs\mu_5^2} \frac{1}{4\wkr\wks} \nn \\ && \hspace{-2.7cm}
	\times \Big[ \mathcal{N}_h(k^0=-\wkr) \fb{1-\fmkr -\fpks +2\fmkr\fpks}\delta(q_0-\wkr-\wks) 
	+ \mathcal{N}_h(k^0=\wkr)  \fb{1-\fpkr -\fmks +2\fpkr\fmks} \delta(q_0+\wkr+\wks) \nn \\ && \hspace{-2.7cm}
	+ \mathcal{N}_h(k^0=\wkr) \fb{-\fpkr -\fpks +2\fpkr\fpks} \delta(q_0+\wkr-\wks) 
	+ \mathcal{N}_h(k^0=-\wkr) \fb{-\fmkr -\fmks +2\fmkr\fmks} \delta(q_0-\wkr+\wks) \Big]_{\bm{q}=\bm{0}}.
	\label{ImPi.2}
\end{eqnarray}
We note that, the remaining $d|\bm{k}|$ integral of Eq.~\eqref{ImPi.2} can be performed using the Dirac delta functions present in the integrand.

\begin{figure}[h]
	\includegraphics[angle=0,scale=0.5]{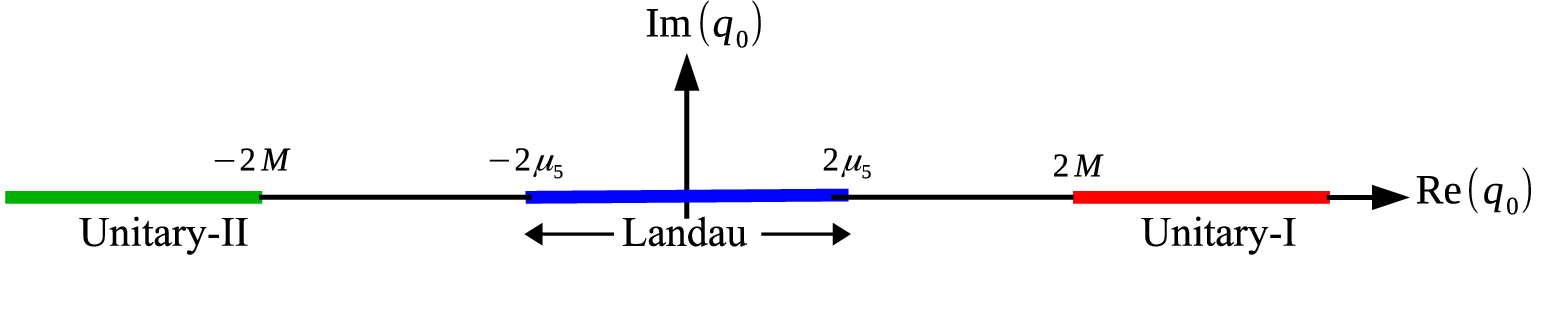} 
	\caption{(Color Online) The analytic structure of the polarization function $\Pi_h(q^0,\bm{q}=\bm{0})$ showing its branch cuts in the complex $q_0$ plane.}
	\label{fig.analytic}
\end{figure}

$\IM\Pi_h(q_0,\bm{q}=\bm{0})$ in Eq.~\eqref{ImPi.2} contains sixteen Dirac delta functions and they give rise to branch cuts of the polarization function in the complex energy plane. The terms with $\delta(q_0-\wkr-\wks)$ and $\delta(q_0+\wkr+\wks)$ are the Unitary-I and Unitary-II cuts respectively, whereas the terms with $\delta(q_0+\wkr-\wks)$ and $\delta(q_0-\wkr+\wks)$ are respectively termed as Landau-I and Landau-II cuts. Each of these Unitary and Landau cuts in turn consists of four sub-cuts corresponding to different helicities $ (r,s) $. Each of the sixteen Dirac delta functions in Eq.~\eqref{ImPi.2} are non-vanishing at different respective kinematic domains which are tabulated in Eq.~\eqref{tab.kin}
\begin{eqnarray}
	\begin{tabular}{|c|c|}
		\hline 
		Dirac Delta Function & \makecell[c]{ ~ \\ Kinematic Regions \\ ~ } \\
		\hline \hline
		$\delta(q_0\mp\wkr\mp\wks)$ & \makecell[c]{ $ 2M \le \pm q_0 < \infty$ ~~if~~ $(r,s)=(-,-)$,\\ 
			$ 2\sqrt{M^2+\mu_5^2} \le \pm q_0 < \infty$ ~~otherwise~~ } \\
		\hline
		$\delta(q_0\pm\wkr\mp\wks)$ & \makecell[c]{ $ 0 \le \pm q_0 \le 2\mu_5$ ~~if~~ $(r,s)=(-,+)$,
			\\ $ -2\mu_5 \le \pm q_0 \le 0$ ~~if~~ $(r,s)=(+,-)$ } \\
		\hline 
	\end{tabular}
	\label{tab.kin}~.
\end{eqnarray}
In Eq.~\eqref{ImPi.2}, when helicity indices $(r,s)$ are summed over, we find from Eq.~\eqref{tab.kin} that, the kinematic domain for Unitary-I cut is $2M\leq^0<\infty$ whereas the same for Unitary-II cut is $-\infty<q^0\le2M$; on the other hand, both the Landau cuts lie in the region $|q^0|\le2\mu_5$. The complete analytic structure of the polarization function $\Pi_h(q^0,\bm{q}=\bm{0})$ has been shown in Fig.~\ref{fig.analytic}. It is evident from Fig.~\ref{fig.analytic} that, for non-zero $\mu_5$ the Landau cuts contribute to the physical time-like region defined in terms of $q^0>0$ and $q^2>0$ which is a purely finite CCP effect. Moreover, for small values of $\mu_5<2M$, there exist a forbidden gap between the Unitary and Landau cuts. However for sufficiently high values of $\mu_5>2M$, this forbidden gap becomes zero which is also a purely finite CCP effect.

Having calculated the real and imaginary parts of the polarization functions $\Pi_h(q;T,\mu_5)$ from Eqs.~\eqref{RePi.2} and \eqref{ImPi.2}, they can be substituted into Eq.~\eqref{Dh} to obtain the meson propagators $D_h$. Using $D_h$, we will in turn calculate the in-medium mesonic spectral function $S_h$ defined as
\begin{eqnarray}
	S_h(q;T,\mu_5) = \frac{1}{2G\pi} \IM D_h(q;T,\mu_5) = \frac{1}{2G\pi} \frac{2G\IM\Pi_h}{(1-2G\RE\Pi_h)^2+(2G\IM\Pi_h)^2}. \label{spec}
\end{eqnarray}
The in-medium mesonic spectral function defined in Eq.~\eqref{spec} contains both the real and imaginary parts of the polarization function and thus it carries all the information about the propagation of the mesonic excitation in the medium. In this work, we also aim to calculate the masses of the mesons in scalar and pseudoscalar channels. However mass can not be uniquely defined corresponding to such collective excitation. For example, the mesonic mass can be defined as the position of the pole of the propagator in the complex energy plane at vanishing three-momentum in which case, the real(imaginary) part of the pole corresponds to the mass(width) of the mesons~\cite{Zhuang:1994dw}; although the problem of finding a complex pole is a bit involved, it can be simplified by considering the small width approximation in which case the equations for the mass and width are decoupled~\cite{Zhuang:1994dw}. Alternatively, one can also define the mesonic mass as the position of the global maximum of the spectral function at vanishing three-momentum. In this work, we will be using a hybrid method to calculate the mass of the mesons as we describe below.

At vanishing three-momenta ($\bm{q}=\bm{0}$), the position of the pole $q^0=M_h$ of the propagator [from Eq.~\eqref{Dh}] 
\begin{eqnarray}
D_h(q^0,\bm{q}=\bm{0};T,\mu_5) = \frac{2G}{1-2G\RE\Pi_h(q^0,\bm{q}=\bm{0};T,\mu_5)-2iG\IM\Pi_h(q^0,\bm{q}=\bm{0};T,\mu_5)}	
\end{eqnarray}
gives the mesonic mass $M_h$ which essentially requires solving the transcendental equation 
\begin{eqnarray}
	1-2G\RE\Pi_h(q^0=M_h,\bm{q}=\bm{0};T,\mu_5)=0 \label{pole}
\end{eqnarray}  
for $M_h$. 
It is easy to realize that, for $\IM\Pi_h(q^0=M_h,\bm{q}=\bm{0};T,\mu_5) \approx 0$ near the pole (small widths), the pole position coincides with the peak position of the spectral function as can be understood from Eq.~\eqref{spec}. It has been observed that at finite CCP Eq.~\eqref{pole} may not admit a real solution for $M_h$. In such a scenario the mesonic mass will be obtained from the the position of global maximum $q^0=M_h$ of the spectral function $S_h(q^0=M_h,\bm{q}=\bm{0};T,\mu_5)$ at vanishing three-momenta ($\bm{q}=\bm{0}$).

%~~~~~~~~~~~~~~~~~~~~~~~~~~~~~~~~~~~~~~~~~~~~~~~~~~~~~~~~~~~~~~~~~~~~~~~~~~~~~~~~~~~~~~~~~~~~~~~~~~~~~~~~~~~~~~~
\section{Numerical Results \& Discussions} \label{sec.results}
Let us first specify the values of the parameters of the NJL model that are used in this work. We have used the NJL coupling $G=5.742$ GeV$^{-2}$, the current quark mass $m=5.6$ MeV, and smooth three-momentum cutoff $\Lambda=568.69$ MeV; these values of the parameters successfully reproduce the experimental/phenomenological vacuum values of pion decay constant $f_\pi\simeq92$ MeV, quark-condensate $\ensembleaverage{\bar{u}u}=\ensembleaverage{\bar{d}d}\simeq -(245)^3$ MeV$^3$, and the pion-mass $m_\pi\simeq140$ MeV. Further, with these choice of the parameters, the dressed quark mass in vacuum comes out to be $M\simeq343.6$ MeV. Although the analytical calculation in Secs.~\ref{sec.gap}-\ref{sec.propagator} have been done keeping the chemical potential $ \mu  $, we plan to give numerical results in this section by taking $\mu=0$ in order to see the interplay between $ \mu_5 $ and $ T $. Numerical results taking all parameter $\{T,\mu,\mu_5\}$ will be reported elsewhere which definitely will be interesting to explore.

A general comment on the use of two flavor NJL model in the present context is in order here. It is well known that, the inclusion of Polyakov loop leads to a better description of the QCD phenomenology, in particular the thermodynamic quantities. However, as highlighted in~\cite{Hansen:2006ee,Chaudhuri:2019lbw,Chaudhuri:2020lga}, incorporating the Polyakov loop does not bring about any qualitative differences in the  variation of the masses of neutral $\pi$ and $\sigma$ mesons as a function temperature. This is attributed to the fact that the constituent quark mass is the key quantity governing the temperature evolution of mesonic masses. Furthermore, the spectral properties of $\sigma$ and neutral $\pi$ mesons are expected to remain qualitatively unchanged even with the inclusion of strange quarks in a three-flavor model~\cite{Rehberg:1995kh,Costa:2008dp}. Instead, use of a two-flavor NJL model facilitates a less cumbersome calculation enabling a focused exploration of the interplay between $T$ and $\mu_5$.
\begin{figure}[h]
	\includegraphics[angle=-90,scale=0.23]{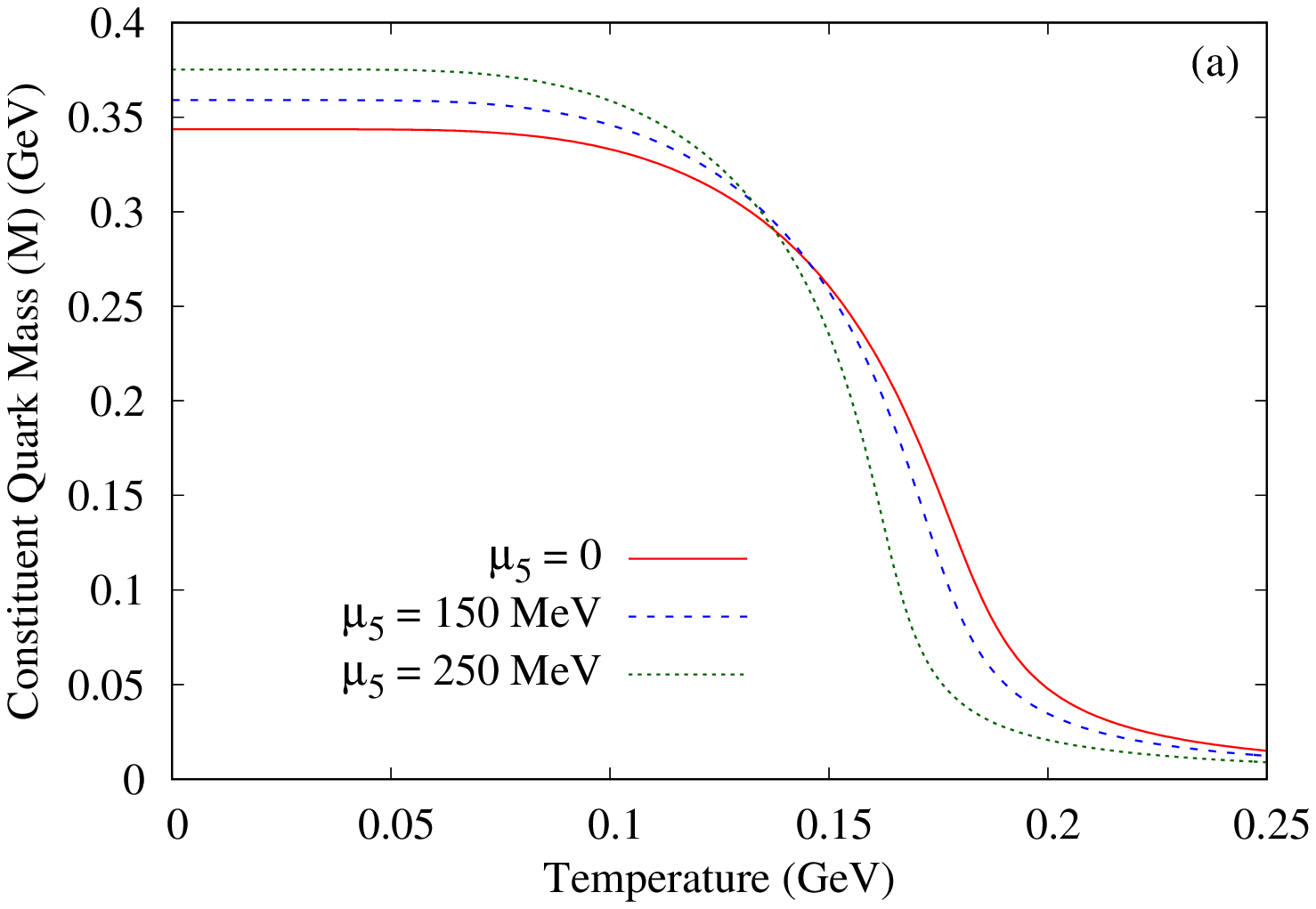} 
	\includegraphics[angle=-90,scale=0.23]{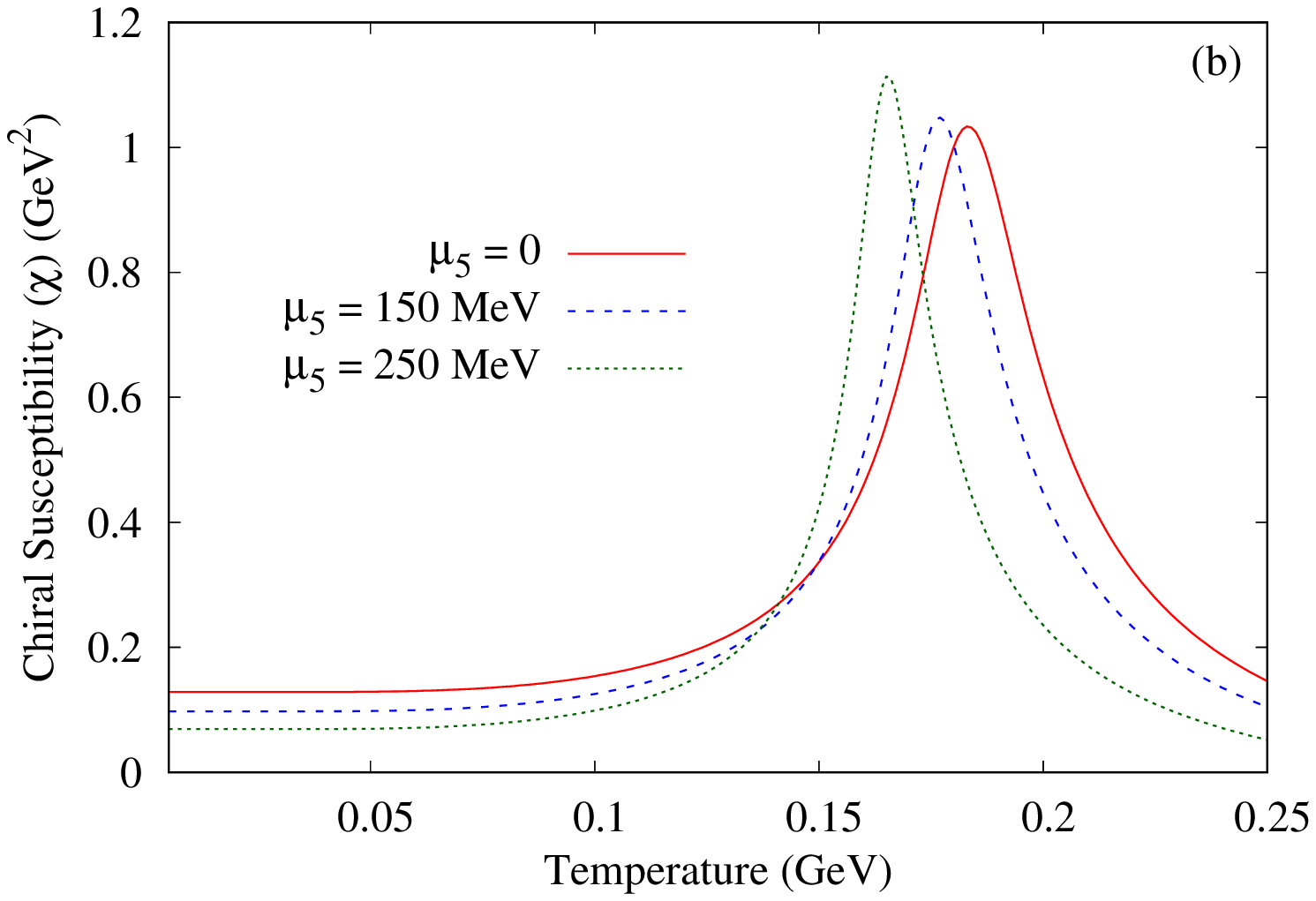} 
	\includegraphics[angle=-90,scale=0.23]{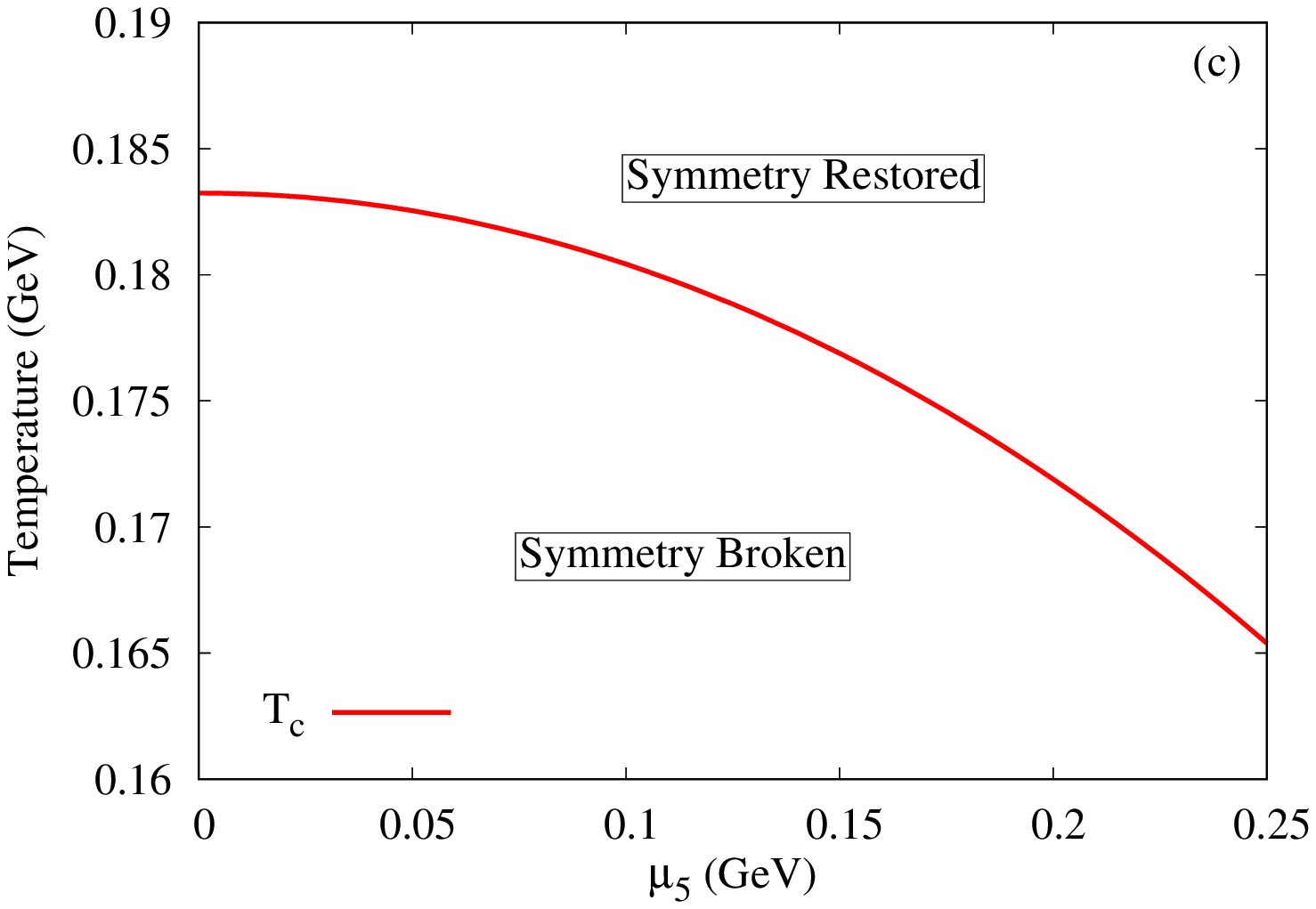} 
	\caption{(Color Online) The variation of the (a) constituent quark mass $M$, (b) chiral susceptibility $\chi$ as a function of temperature $T$ at different values of CCP. (c) The chiral phase diagram on the $T-\mu_5$ plane showing the variation of the pseudo-chiral phase transition temperature $T_c$ as a function of CCP.}
	\label{fig.M}
\end{figure}

The constituent quark mass $M$ has been obtained by solving the gap equation given by Eq.~\eqref{gap.final}. As can be seen in Fig.~\ref{fig.M}(a), $M$ is large ($\sim 300$ MeV) in the low temperature region (irrespective of the value of $\mu_5$) due to the spontaneous breaking of the chiral symmetry characterized by the large values of the quark condensate. As the temperature increases, $M$ initially remains constant up to a certain temperature. Beyond this, $M$ gradually decreases, eventually reaching the values of the bare quark mass. This behaviour is a consequence of the pseudo-chiral phase transition. In the high temperature region, irrespective of the values of CCP, the constituent quark mass $M$ asymptotically approaches to the current quark mass $m$ value. If we compare the curves for different CCP of Fig.~\ref{fig.M}(a), we notice that, with the increase in CCP, $M$ also increases in the low temperature region. Moreover, the transition temperature of the pseudo-chiral phase transition decreases with the increase in CCP i.e. and the sudden drop of the constituent quark mass occurs at relatively smaller temperature for higher values of CCP. Hence, CCP has the tendency to make the chiral condensate stronger at low temperature region indicating a ``chiral catalysis'' (CC) where the chiral imbalance (characterized by $\mu_5$) catalyzes the dynamical symmetry breaking. On the other hand, an opposite effect is observed on the high temperature region in which the $\mu_5$ weakens the chiral condensate so that chiral symmetry is restored at a relatively lower temperature as compared to the $\mu_5=0$ case; thus indicating an ``inverse chiral catalysis'' (ICC) in which the dynamical symmetry breaking is opposed by the presence of a chiral imbalance~\cite{Ghosh:2022xbf}. 

Having obtained the constituent quark mass, we have now calculated the chiral susceptibility 
\begin{eqnarray}
	\chi = \frac{1}{2G}\FB{\frac{\del M}{\del m}-1}
\end{eqnarray}
which is plotted as a function of temperature at different CCP in Fig.~\ref{fig.M}(b). The position of the maximum of the susceptibility corresponds to the pseuodo-chiral phase transition temperature $T_c$~\cite{Zhao:2006br,Chaudhuri:2019lbw}. As can be observed in Fig.~\ref{fig.M}(b), with the increase in CCP, the position of the maximum moves towards lower values of temperature indicating an ICC as described in the previous paragraph. The explicit variation of $T_c$ as a function of CCP has been depicted in Fig.~\ref{fig.M}(c) from which we notice that $T_c$ decreases monotonically with the increase in $\mu_5$ owing to the ICC.
%~~~~~~~~~~~~~~~~~~~~~~~~~~~~~~~~~~~~~~~~~~~~~~~~~~~~~~~~~~~~~~~~~~~~~~~~~~~~~~~~~~~~~~~~~~~~~~~~~
%
\subsection{Mesonic Properties}
\begin{figure}[h]
	\includegraphics[angle=-90,scale=0.45]{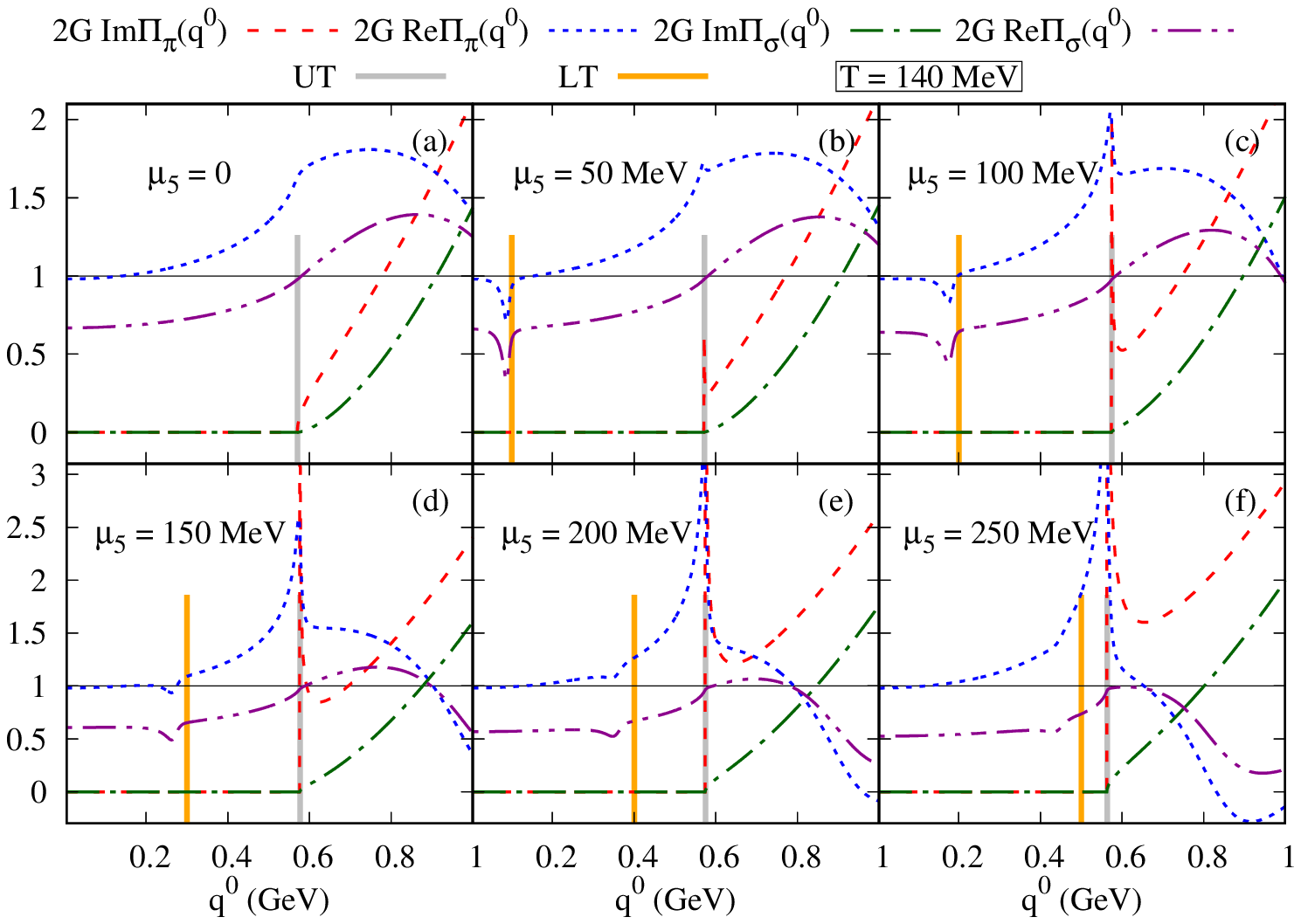} 
	\includegraphics[angle=-90,scale=0.45]{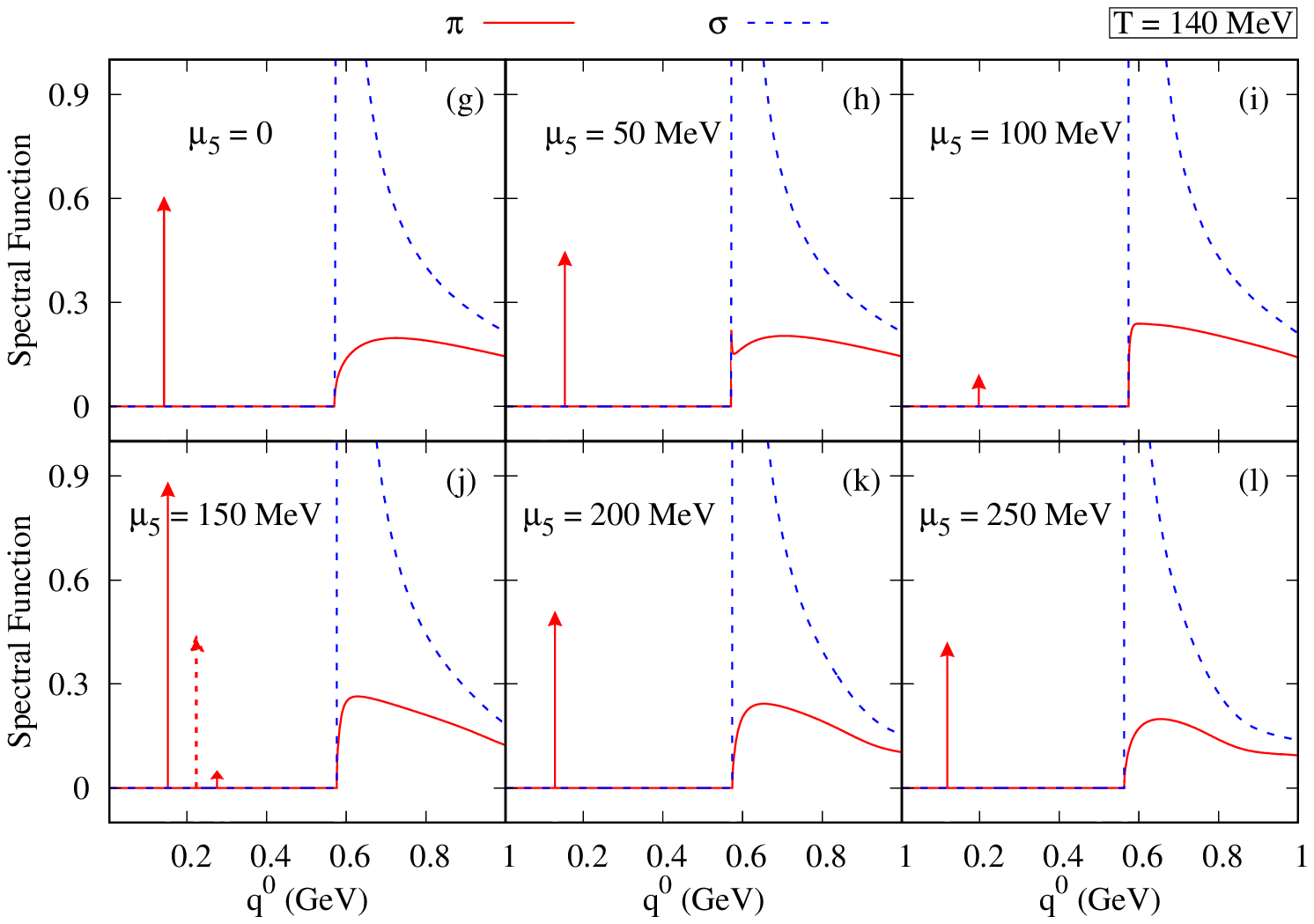} 
	\caption{(Color Online) (a)$ - $(f) The scaled polarization functions and (g)$ - $(l) spectral functions of $\pi$ and $\sigma$ as a function of $q^0$ for different values of CCP ($\mu_5=0, 50, 100, 150, 200$ and 250 MeV) at $T=140$ MeV. `UT' and `LT' denotes the Unitary and Landau cut thresholds on the $q^0$ axis. The arrows represent the Dirac delta functions corresponding to the poles of the propagator; the length of the arrows are scaled by the value of the residues at the poles and the dashed arrows correspond to negative residues.}
	\label{fig.pola.140}
\end{figure}

\begin{figure}[h]
	\includegraphics[angle=-90,scale=0.45]{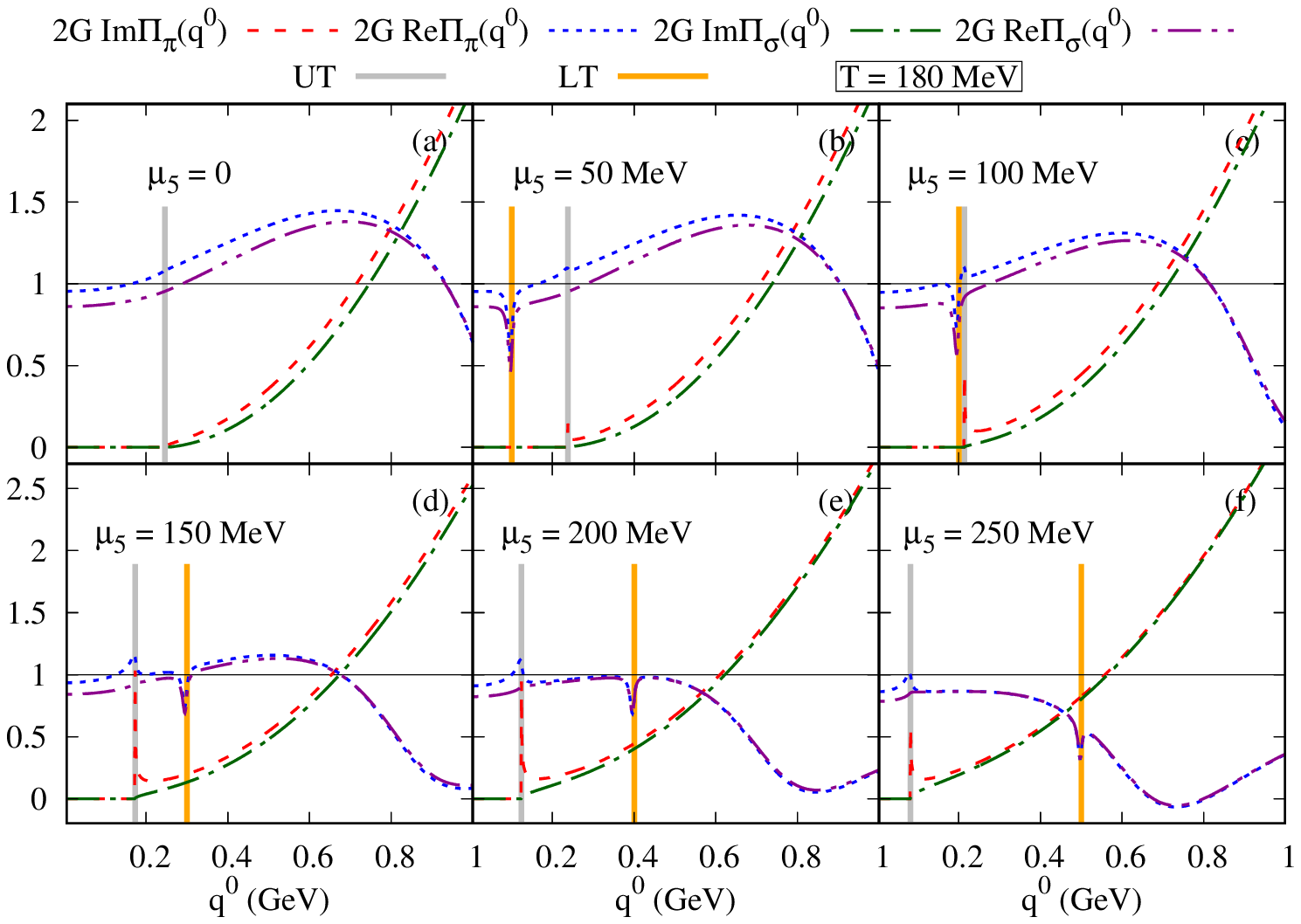} 
	\includegraphics[angle=-90,scale=0.45]{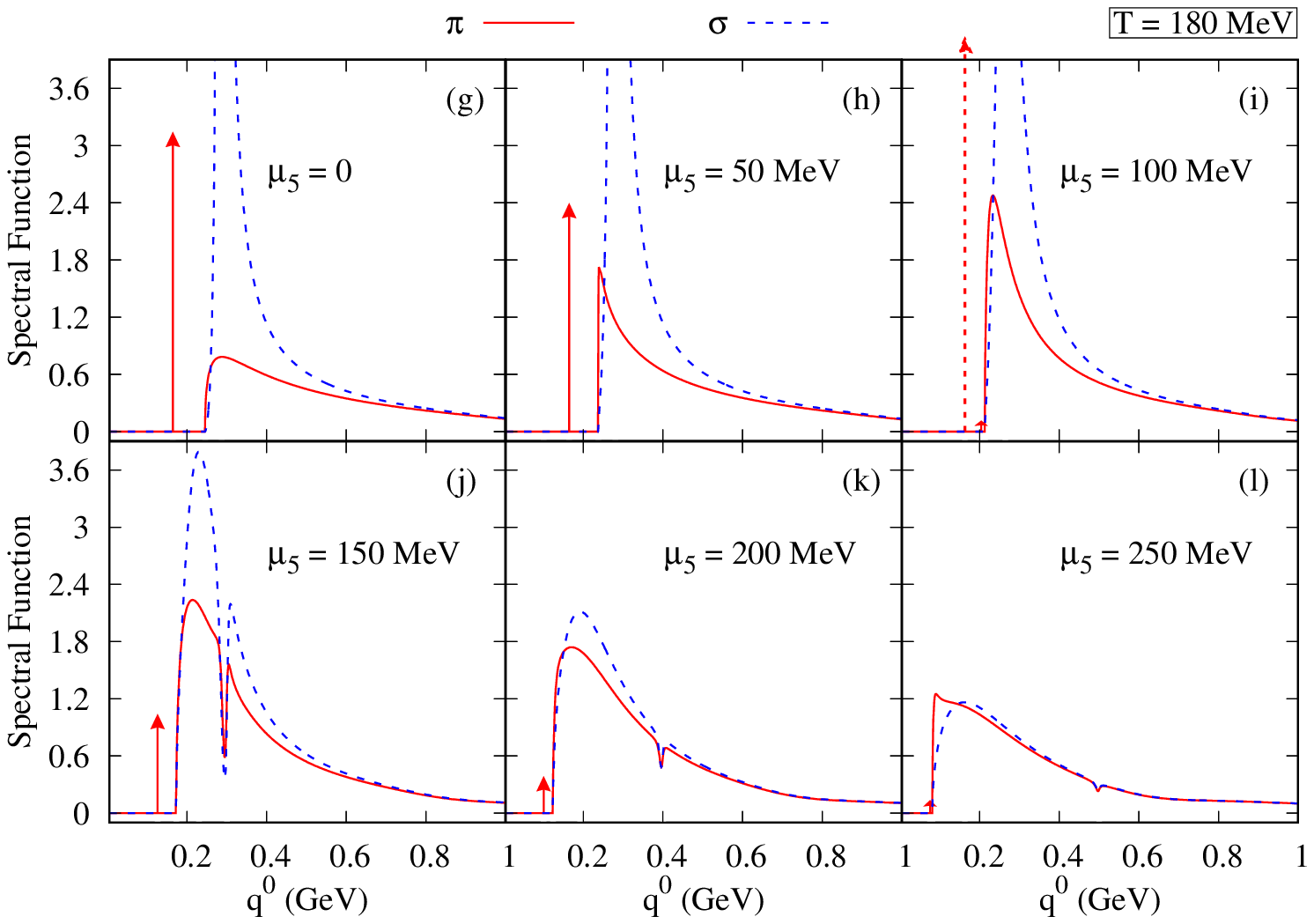} 
	\caption{(Color Online) (a)$ - $(f) The scaled polarization functions and (g)$ - $(l) spectral functions of $\pi$ and $\sigma$ as a function of $q^0$ for different values of CCP ($\mu_5=0, 50, 100, 150, 200$ and 250 MeV) in the vicinity of pseudo-chiral transition temperature. `UT' and `LT' denotes the Unitary and Landau cut thresholds on the $q^0$ axis. The arrows represent the Dirac delta functions corresponding to the poles of the propagator; the length of the arrows are scaled by the value of the residues at the poles and the dashed arrows correspond to negative residues.}
	\label{fig.pola.180}
\end{figure}

\begin{figure}[h]
	\includegraphics[angle=-90,scale=0.45]{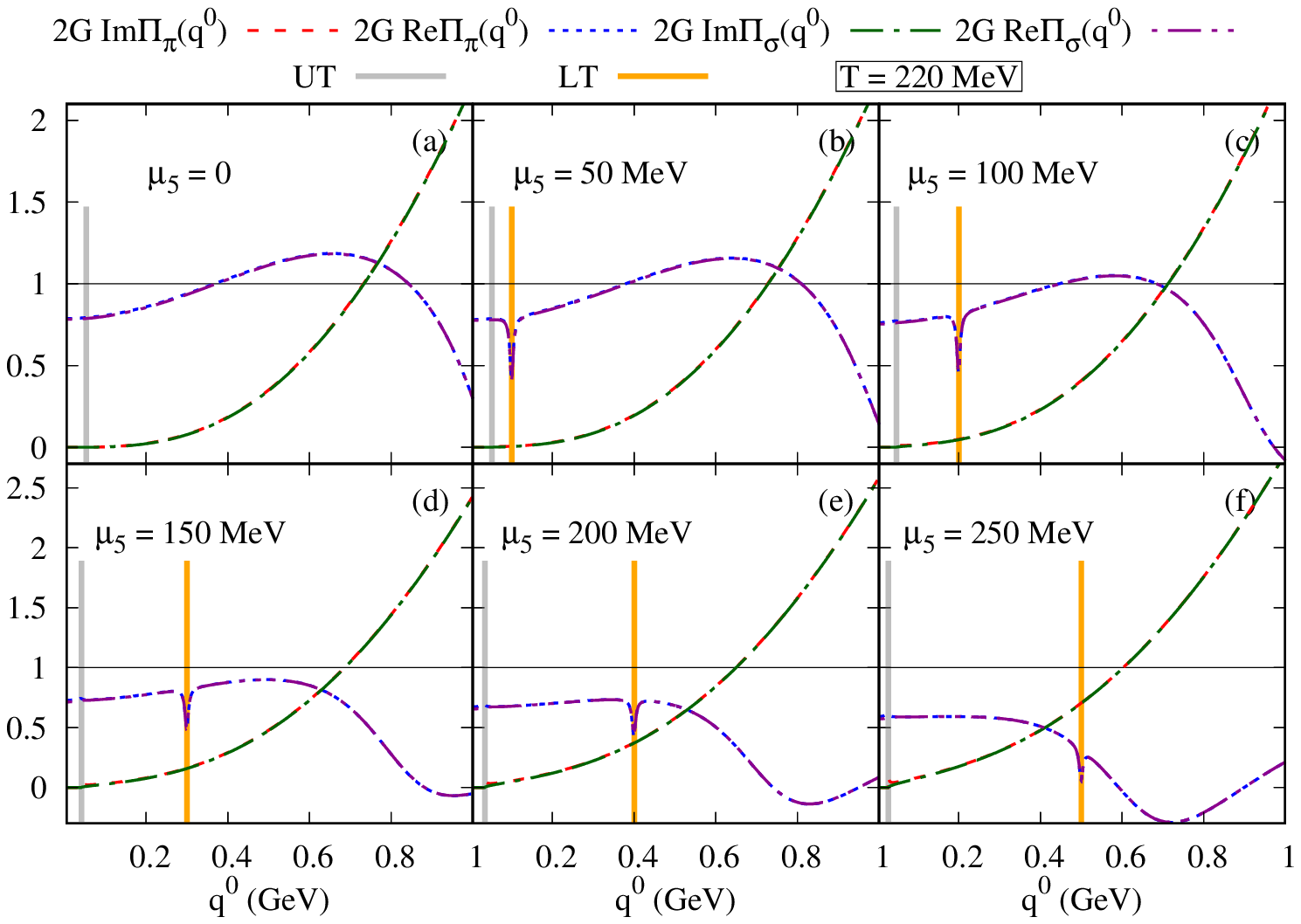} 
	\includegraphics[angle=-90,scale=0.45]{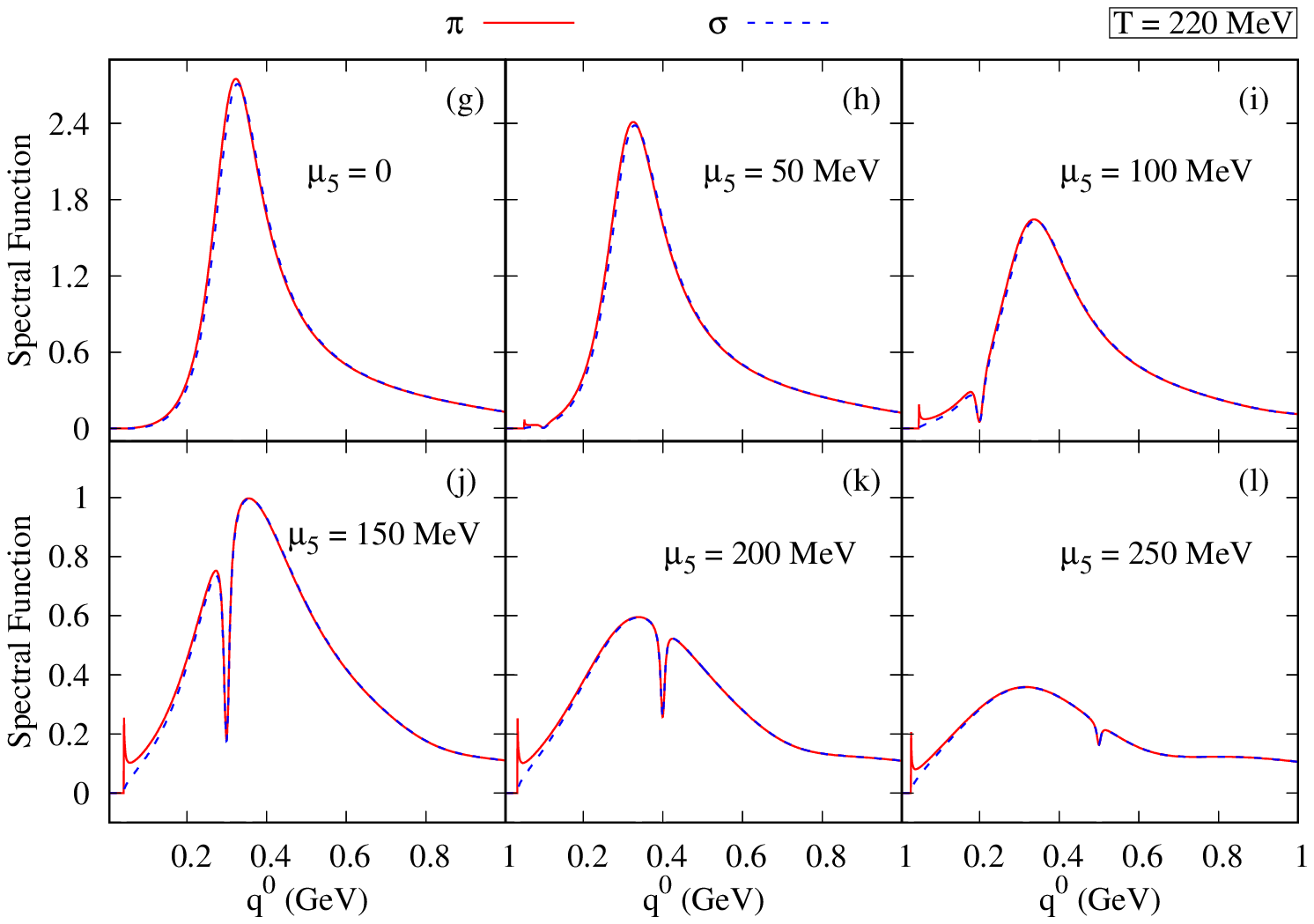} 
	\caption{(Color Online) (a)$ - $(f) The scaled polarization functions and (g)$ - $(l) spectral functions of $\pi$ and $\sigma$ as a function of $q^0$ for different values of CCP ($\mu_5=0, 50, 100, 150, 200$ and 250 MeV) at $T=220$ MeV. `UT' and `LT' denotes the Unitary and Landau cut thresholds on the $q^0$ axis.}
	\label{fig.pola.220}
\end{figure}

Let us now move on to the numerical results of the mesonic properties. 
In Figs.~\ref{fig.pola.140}$ - $\ref{fig.pola.220} of this subsection, we will study the polarization functions and the spectral properties of $ \pi $  and $ \sigma $ meson at three different values of temperature capturing different stages of chiral phase transition. These are as follows:
\begin{enumerate}
	\item $ T<  T_{0}^{ \rm~ch}$ : his scenario represents a situation where chiral symmetry is broken. Upon analyzing Figs.~\ref{fig.M}(a), we choose $ T = 140 $~MeV to examine this case.
	\item $ T\approx   T_{0}^{ \rm~ch}$ : Here, we select $ T = 180 $ MeV to investigate properties in the vicinity of the chiral phase transition.
	\item $ T >  T_{0}^{ \rm~ch}$ : We will consider $ T = 220 $ MeV, which corresponds to the (partial) restoration of chiral symmetry.
\end{enumerate}

In Figs.~\ref{fig.pola.140}(a)$ - $(f), we have depicted the variations of the dimensionless quantities $2G\RE\Pi_h(q^0,\bm{q}=\bm{0})$ and $2G\IM\Pi_h(q^0,\bm{q}=\bm{0})$ as a function of $q^0$ for six different values of CCP (with $\mu_5=0$, 50, 100, 150, 200, and 250 MeV) in the chiral symmetry broken phase. Here the Landau cut (at $q^0=2\mu_5$) and the Unitary cut (at $q^0=2M$) thresholds are indicated by orange and grey colored vertical lines respectively. 
Let us first describe the variation of $\IM\Pi_h(q^0,\bm{q}=\bm{0})$ as a function of $q^0$ in the physical time-like region $q^0>0$ for different values of CCP. At vanishing CCP, the $\IM\Pi_h(q^0,\bm{q}=\bm{0})$'s do not have contributions from the Landau cut in the kinematic region $q^0>0$, so these curves start from the Unitary cut threshold $q^0=2M$ as can be noticed from the red-dash and green-dash-dot curves in the Fig~\ref{fig.pola.140}(a). In Figs.~\ref{fig.pola.140}(a)$ - $(f), as we increase the value of CCP (from zero to 250 MeV in steps of 50 MeV), the unitary cut thresholds (which is $q^0=2M$) are found to move towards higher values of $q^0$ by a small amount ($\sim$ by few tens of MeVs). This is due to the increase in $M$ with the increase in $\mu_5$ owing to the CC as can be understood from Fig.~\ref{fig.M}(a) by comparing its red-solid, blue-dash and green-dot curves at temperature slice $T=120$ MeV.

At non-zero values of $\mu_5$, the $\IM\Pi_h(q^0,\bm{q}=\bm{0})$'s have contributions from the Landau cuts as well. However the Landau cut contributions have sub-leading magnitudes as compared to  the Unitary cuts, thus making them invisible for the ordinate ranges considered in the graphs. Moreover, the Landau cut contributions are so small that, they have negligible contribution on the numerical results of mesonic spectral functions and masses and henceforth we will not discuss them in the rest of this section.
Concentrating on Figs.~\ref{fig.pola.140}(a)$ - $(f), it can be seen that at finite values of $\mu_5$, the $\IM\Pi_\pi(q^0,\bm{q}=\bm{0})$'s (corresponding to the pion) become non-monotonic near the Unitary cut threshold; in particular, as we increase the value of $q^0$, $\IM\Pi_\pi(q^0,\bm{q}=\bm{0})$'s at first start to increase at the Unitary cut threshold to a local maxima, then it decreases to a local minima, and finally it again increases monotonically at high values of $q^0$. On the other hand, $\IM\Pi_\sigma(q^0,\bm{q}=\bm{0})$'s (corresponding to the sigma) increase monotonically after the Unitary cut threshold. Unlike $\IM\Pi_\pi(q^0,\bm{q}=\bm{0})$, a non-monotonic behaviour is observed in $\RE\Pi_h(q^0,\bm{q}=\bm{0})$ for both $ \pi $ and $ \sigma $. Near each of the Landau and Unitary cut threshold, $\RE\Pi_h(q^0,\bm{q}=\bm{0})$'s suffer a sudden change in its curvature owing to the analyticity of the polarization function.  Now, to evaluate the masses of $ \pi $ and $ \sigma $ one has to solve Eq.~\eqref{pole}. Therefore, the intersection point of $2G\RE\Pi_h(q^0,\bm{q}=\bm{0})$ and the horizontal line at unity (indicated by the black solid line) determines the solution of Eq.~\eqref{pole}, providing the mass of the mesons. Upon closer examination of Fig.~\ref{fig.pola.140}(d) at $\mu_5=150$ MeV, we observe that the blue curve intersects with the line of unity three times indicating the presence of multiple solutions (three) for the pion mass. However, as the value of CCP increases (to $\geq 200$ MeV), there is only one intersection point between the blue and black curves, indicating a single solution to Eq.~\eqref{pole}. Hence, this phenomenon of multiple solutions to Eq.~\eqref{pole}, or the multi-valued nature of the pion mass, occurs in a small region of the $\mu_5$ parameter space near $\mu_5 = 150$ MeV at a temperature of 120 MeV. It is important to note that this effect is solely due to finite CCP and does not have a clear physical interpretation. However, no such feature is observed in case of mass of the $ \sigma  $ meson as it remains single valued for all values of $ \mu_5 $.
It should be noted that, for a high value of $ q^0 $, independent of the values of $ \mu_5 $, for both $ \pi $ and $ \sigma $ meson $2G \RE\Pi_h(q^0,\bm{q}=\bm{0})$ intersects the black line at unity which in principle corresponds to another solutions for both $ \pi $ and $ \sigma $ like modes/excitations. 
However these solutions at large values of $q^0$ are unphysical since these poles have negative residues~\cite{Oertel:1999fk}. The residue $R(M_h)$ at a pole $q^0=M_h$ in the small width approximation~\cite{Klevansky:1992qe} is given by the expression $R(M_h)=-\FB{\frac{\del\RE\Pi_h}{\del q_0^2}}^{-1}\Big|_{q_0=M_h}$. The residue is proportional to the square of the quark-meson coupling and thus provides an estimate of the strength of interaction. A negative residue at the pole corresponds to unphysical mode owing to an imaginary pion-quark coupling constant~\cite{Oertel:1999fk}. Hence, those solutions are neglected throughout this article.

In Figs~\ref{fig.pola.140}(g)$ - $(l), we have illustrated the spectral functions of $ \pi $ and $ \sigma $ mesons as functions of $q^0$ for six different values of CCP (with $\mu_5=0$, 50, 100, 150, 200, and 250 MeV) when chiral symmetry is spontaneously broken. In all these plots, an arrow is used to indicate the Dirac delta function which corresponds to the pole of the propagator. In Figs~\ref{fig.pola.140}(g)$ - $(l), we have scaled the length of the arrows (Dirac delta) by the value of the residues at the poles and the dashed arrows in the figure correspond to a negative residue.
At vanishing CCP, the pion spectral function consists of a Dirac delta function at its pole, along with a continuum from the two-quark threshold. In contrast, the sigma spectral function shows a Breit-Wigner-like structure starting from the two-quark threshold. This is a manifestation of the fact that in the low-temperature region, pions are bound states of $q\bar{q}$, whereas sigmas are resonant states with a finite decay width, which is proportional to the width of the Breit-Wigner structure.
As the CCP increases up to 100 MeV, the position of the Dirac delta function moves towards higher values of $q^0$ and the value of residue decreases, which is consistent with the shift in the intersection point of the blue and black curves observed in Figs.~\ref{fig.pola.140}(a)$ - $(c). On the other hand, the continuum threshold for the pions, as well as the threshold of the sigma spectral function, both move towards higher values of $q^0$. These shifts are a consequence of the movement of the Unitary cut threshold.
At $\mu_5=150$ MeV, the pion spectral function exhibits three Dirac delta functions separated from each other. Out of the three delta functions, the first one exhibits large and positive residue comparable to the $\mu_5=0$ case. Interestingly, the second delta function corresponds to a negative residue whereas the third one has positive but very small residue as compared to the $\mu_5=0$ case. This observation is compatible with the behaviour depicted in Fig.~\ref{fig.pola.140} (d), which shows three solutions for the pion mass, as discussed previously.
Moreover, at higher values of CCP ($\ge200$ MeV), only a single Dirac delta function appears in the pion spectral function. This is in line with the behavior shown in Fig.~\ref{fig.pola.140} (e)$ - $(f).
Furthermore, it should be noted that with the increase in CCP, the width of the sigma meson decreases, making it more stable.

In Figs.~\ref{fig.pola.180}(a)$ - $(f), we have presented the variations of the dimensionless quantities $2G\RE\Pi_h(q^0,\bm{q}=\bm{0})$ and $2G\IM\Pi_h(q^0,\bm{q}=\bm{0})$ as functions of $q^0$ for six different values of CCP (i.e. $\mu_5=0$, 50, 100, 150, 200, and 250 MeV) near the pseudo-chiral transition temperature. Here also the Landau cut (at $q^0=2\mu_5$) and the Unitary cut (at $q^0=2M$) thresholds are indicated by orange and grey colored vertical lines, respectively.
Firstly, we observe that the overall magnitudes of $\IM\Pi_h(q^0,\bm{q}=\bm{0})$ increases for all values of $ \mu_5 $  compared to the scenario when chiral symmetry is spontaneously broken owing to the enhancement of available thermal phase space.
Secondly, it is evident that the quantities $2G\RE\Pi_h(q^0,\bm{q}=\bm{0})$ and $2G\IM\Pi_h(q^0,\bm{q}=\bm{0})$ for both  pion and sigma approach each other owing to the the partial restoration of chiral symmetry. Moreover with the increase in $\mu_5$, the transition temperature of the pseudo-chiral phase transition decreases owing to the ICC effect as can be seen explicitly in Fig.~\ref{fig.M}(c); so that even at $T=180$ MeV, with sufficient high value of $\mu_5 \gtrsim 200$ MeV, the polarization functions of $\pi$ and $\sigma$ almost coincide with each other as a consequence of the partial restoration of chiral symmetry.
Now, based on the discussion of the analytic structure of the imaginary part of the polarization function and the kinematic domain defined in Eq.~\eqref{tab.kin}, we can observe that the threshold for the Landau cut contribution moves towards higher values of $q^0$ with an increase in $\mu_5$. However, the Unitary cut threshold is proportional to $M$. Since the magnitude of $M$ decreases with increasing $\mu_5$ in the vicinity of the pseudo-chiral phase transition due to the ICC effect, in this case, we observe that the Unitary and Landau cut contributions approach each other as we increase $\mu_5$. They almost coincide at $\mu_5\sim 100$~MeV. Further increasing $\mu_5$, the Unitary cut threshold occurs at smaller values of $q^0$ compared to the $q^0$ values for the Landau cut contribution starts.
These observations directly modify the spectral properties of $ \pi $ and $ \sigma $ mesons in chiral imbalanced medium at temperatures close to the pseudo-chiral transition temperature, as shown in Figs.~\ref{fig.pola.180}(g)$ - $(l).  
 Here as well, the pion spectral function consists of a Dirac delta function at its pole, along with a continuum originating from the two-quark threshold and the sigma spectral function exhibits a Breit-Wigner-like structure starting from the two-quark threshold. However, unlike the low temperature case, here the two quark threshold moves towards the lower values of $ q^0 $ with increasing $ \mu_5 $ indicating the ICC effect observed in Fig.~\ref{fig.M}(a). This further leads to the decrease in the gap between the Dirac delta spike and the continuum structure of pion spectral function. 
 Moreover, in this case we observe multiple (two) roots  of pions at $ \mu_5 = 100 $~MeV. One of the pole correspond to a large negative residue whereas the other one has very small positive residue as compared to the $\mu_5=0$ case. Although, $2G\RE\Pi_\pi(q^0,\bm{q}=\bm{0})$ intersects the line of unity (see Fig.~\ref{fig.pola.180} (d)) multiple times but that occurs for $ q^0 $ values greater than the Unitary cut thresholds. Hence they correspond to two quark continuum and we have only one pole for pion at $ \mu_5 = 150 $~MeV. 
 Furthermore, for $ \mu_5\ge 200 $~MeV, both the spectral functions for pion and sigma are almost on top of each other, indicating the partial restoration of chiral symmetry due to the ICC effect. At these high values of CCP, only one Dirac delta with small positive residue appears in the spectral function.

Next we present dimensionless polarization function (both real and imaginary part) and the spectral functions of pion and sigma when chiral symmetry is partially restored for different values of $ \mu_5  $ in left and right panel of Fig.~\ref{fig.pola.220} respectively. All the graphs are labelled using the same scheme as done in Figs.~\ref{fig.pola.140} and \ref{fig.pola.180}. The imaginary part of the polarization functions for both the mesons appears to be thermally enhanced compared to the previous two cases. 
Moreover, the polarization functions (both the real and imaginary parts) of pion and sigma become degenerate due to the partial restoration of chiral symmetry. 
Since the constituent mass of quarks are nearly equal to the bare mass limit, the Unitary cut threshold moves towards very small values of $ q^0 $. As a result, no pole mass is found for pions, which is evident from the plots of spectral function shown in Figs.~\ref{fig.pola.220}(g)$ - $(l). In this case, both pion and sigma appear as resonant state and their spectral functions coincide with each other indicating partial chiral symmetry restoration. Thus we get the so call Mott transition~\cite{MOTT:1968tyq,Hufner:1996pq,Hansen:2006ee} as the pion spectral function has lost its Dirac delta structure and melted to a Breit-Wigner structure similar to the sigma. 
Moreover, from Figs.~\ref{fig.pola.220}(g)$ - $(l), it is evident that with the increase in CCP, the width of the spectral function increases, and the peak decreases. Physically, this corresponds to the enhancement of the decay process of $ \pi $ and $ \sigma $ mesons, implying that they become more unstable for higher values of $ \mu_5 $.
Finally, it should be noted that at high values of CCP, the pion and sigma spectral functions have multiple maxima. This corresponds to the sudden jump observed in the masses of pion and sigma meson in the subsequent figures.
\begin{figure}[h]
	\includegraphics[angle=-90,scale=0.5]{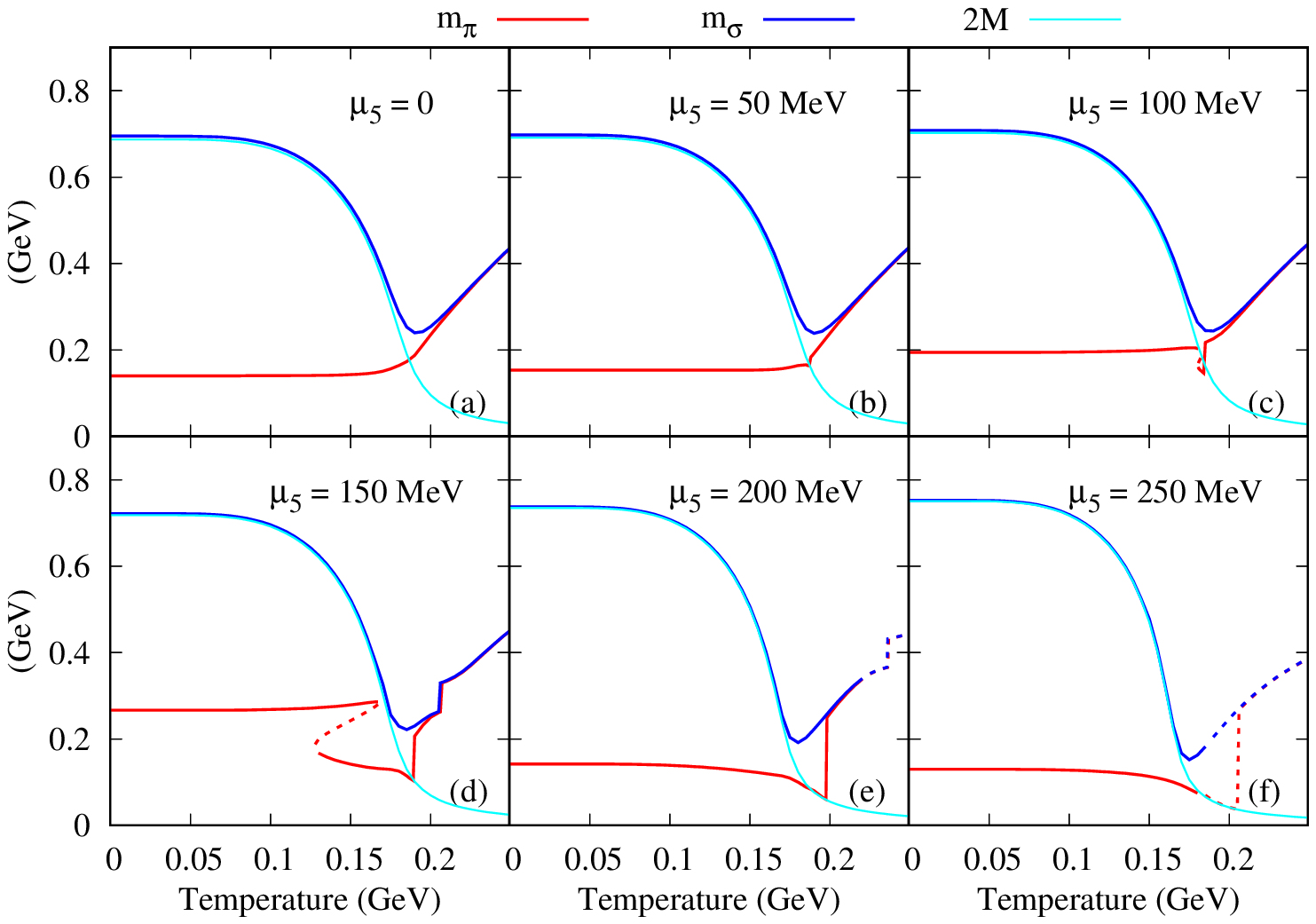} 
	\includegraphics[angle=-90,scale=0.5]{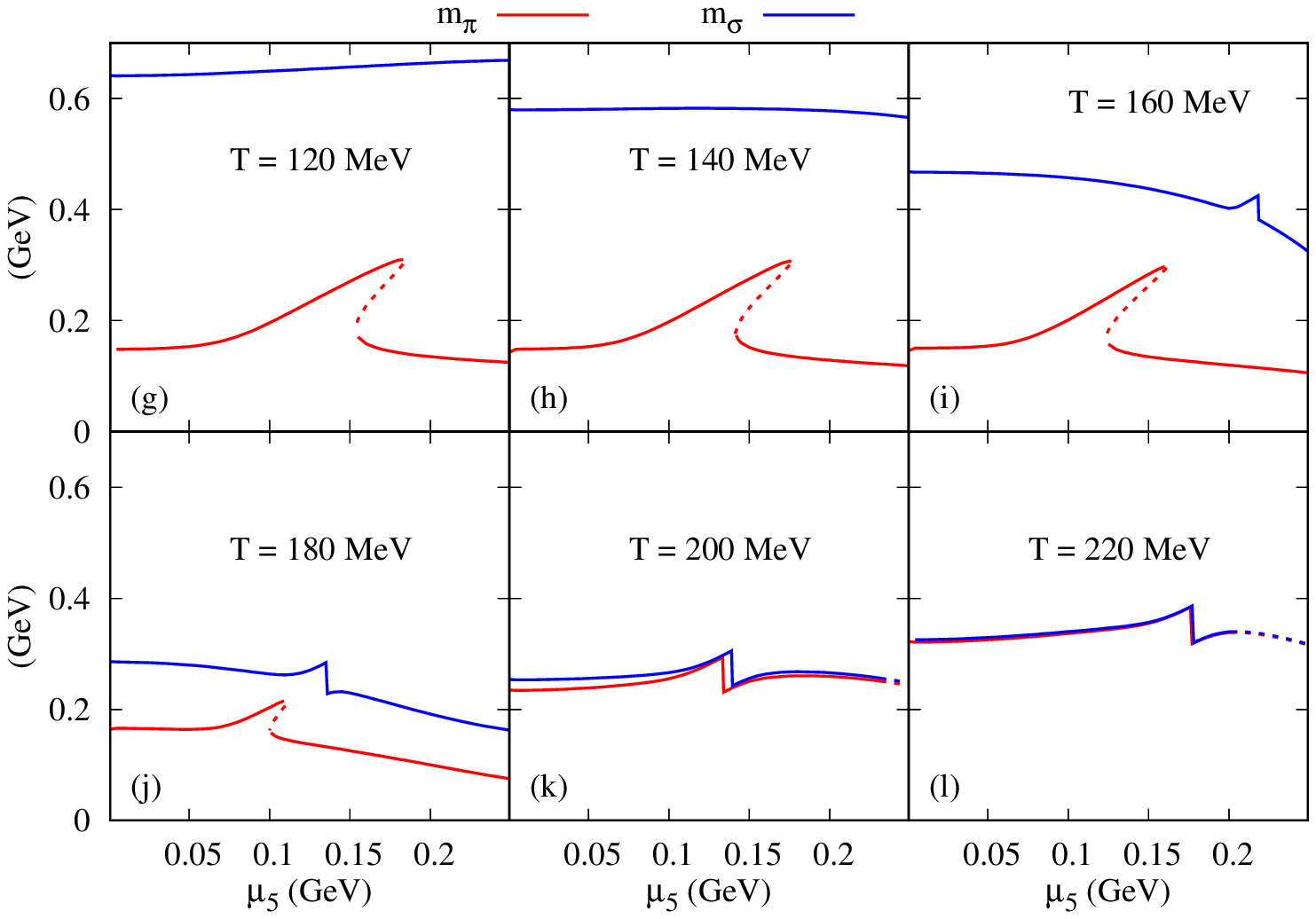} 
	\caption{(Color Online) The masses of $\pi$ and $\sigma$ as a function of (a)$ - $(f) temperature for different values of CCP, and (g)$ - $(l) CCP for different values of temperature. The dashed lines correspond to negative residues at the pole masses. The variation of the twice of the constituent quark mass ($2M$) as a function of temperature is also shown in (a)-(f) by cyan color.}
	\label{fig.mh}
\end{figure}
\begin{figure}[h]
	\includegraphics[angle=-90,scale=0.23]{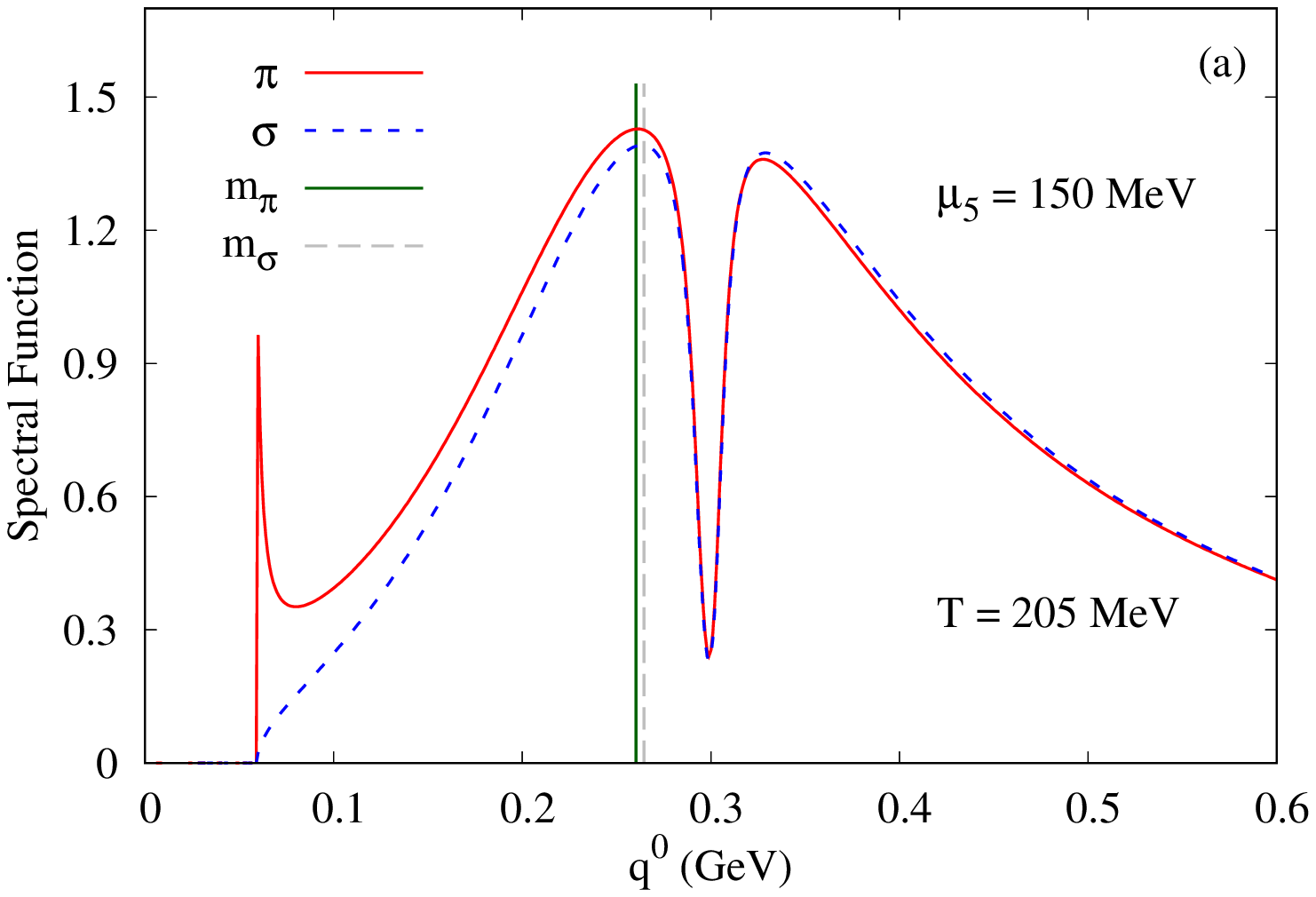} 
	\includegraphics[angle=-90,scale=0.23]{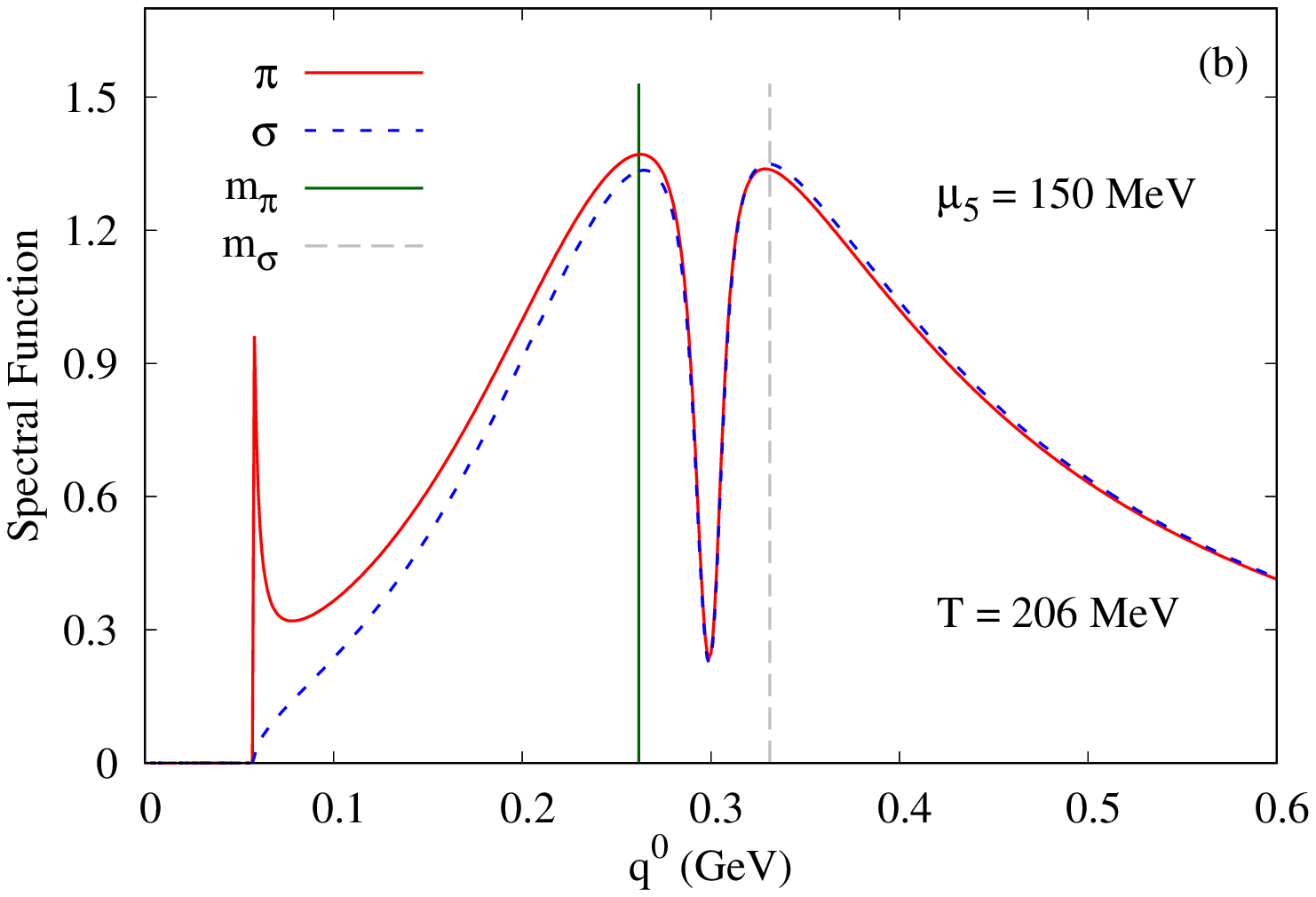}
	\includegraphics[angle=-90,scale=0.23]{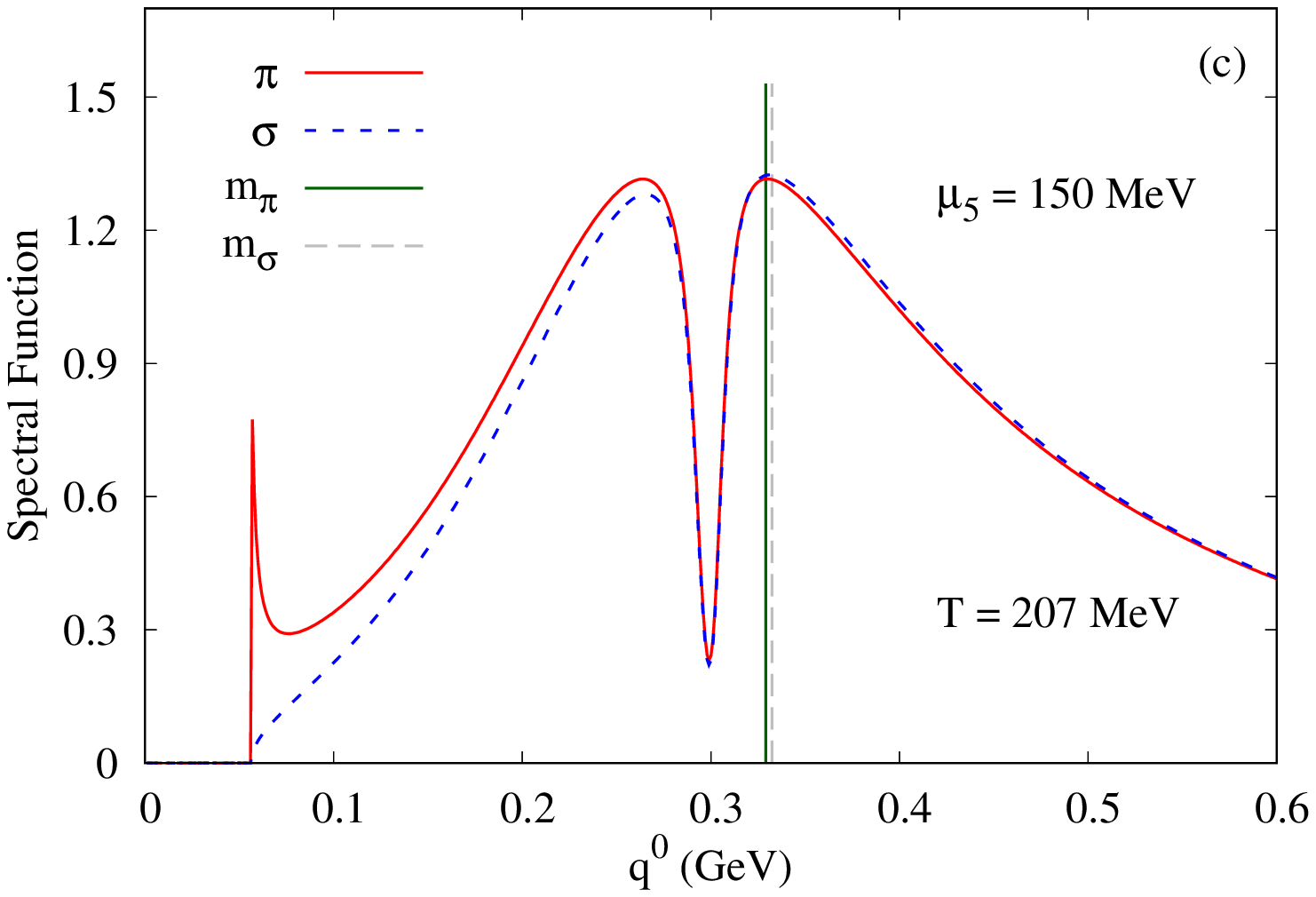}
	\caption{(Color Online) The spectral functions of $\pi$ and $\sigma$ as a function of $q^0$ for $\mu_5=150$ MeV at (a) $T = 205$ MeV, (b) $T = 206$ MeV, and (c) $T = 207$ MeV. The green and grey vertical lines respectively shows the position of $m_\pi$ and $m_\sigma$ on the $q^0$-axis. }
	\label{fig.jump}
\end{figure}
\begin{figure}[h]
	\includegraphics[angle=-90,scale=0.33]{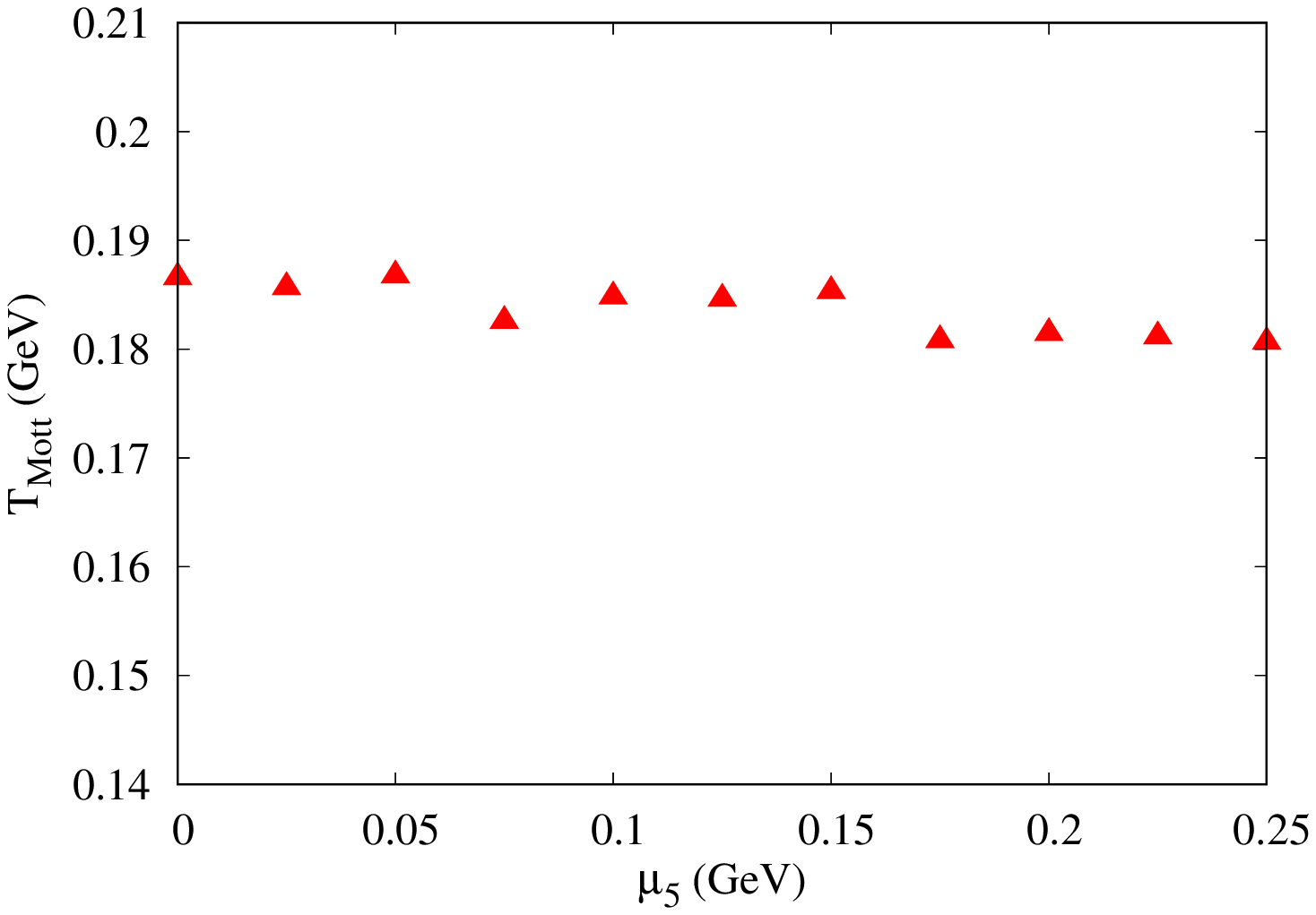} 
	\caption{(Color Online) The variation of the Mott transition temperature $T_\text{Mott}$ as a function of CCP.}
	\label{fig.mott}
\end{figure}

We now shift our focus towards examining how the masses of the $\pi$ and $\sigma$ mesons depend on the variables $T$ and $\mu_5$. In Figs.~\ref{fig.mh}(a)$ - $(f), we have presented the variations of pion and sigma masses as functions of temperature for six different values of CCP (i.e. $\mu_5=0$, 50, 100, 150, 200, and 250 MeV). To obtain these results we have solved Eq.~\eqref{pole} self-consistently. Essentially these are the intersections of blue dotted and violet dot-dot-dashed curves with the black solid line represents unity in left panels of Figs~\ref{fig.pola.140}$ - $\ref{fig.pola.220}. If Eq.~\eqref{pole} lacks real solutions, we have considered the location of the global maxima in the spectral function as an estimate for the mesonic mass. In Figs.~\ref{fig.mh}(a)$ - $(f), we have also plotted the variation of the twice of the constituent quark mass ($2M$) as a function of temperature in cyan color. As evident from the plots, in all the instances, at higher values of temperature mass of pion and sigma mesons coincides with each other indicating partial restoration of chiral symmetry. We also notice that, $M_\sigma$ is always greater than $2M$ for all temperatures and CCP values implying that $\sigma$ meson is always a resonant state whereas $M_\pi$ is larger than $2M$ in the low temperature region only where the pion is a bound state. At sufficiently high value of temperature, $M_\pi$ becomes more than $2M$ making the $\pi$ meson a resonant state indicating a Mott transition~\cite{MOTT:1968tyq,Hufner:1996pq,Hansen:2006ee}. The temperature at which the pion becomes heavier than $2M$ is called the Mott transition temperature and denoted by $T_\text{Mott}$. Therefore, in Figs.~\ref{fig.mh}(a)$ - $(f), the intersections of the red and cyan curves gives $T_\text{Mott}$ i.e. $M_\pi(T\geq T_\text{Mott}) = 2M(T\geq T_\text{Mott})$.
The variation of mesonic mass with respect to $\mu_5$ for different temperature values is illustrated in Figs.~\ref{fig.mh}(g)$ - $(l). Here also it can be seen that, for higher values $ T$, the masses of $ \pi $ and $ \sigma  $ remain degenerate across the entire range of $\mu_5$, suggesting the scenario when chiral symmetry is partially restored. 
 Moreover, from the figures presented in both the left and right panels of Fig.~\ref{fig.mh}, it becomes apparent that the masses of both $\pi$ and $\sigma$ mesons exhibit highly non-monotonic behaviour with respect to temperature and CCP. Furthermore,  the pion mass demonstrates a multivalued behaviour within the intermediate range of $T$ and $\mu_5$. This characteristic can be explained by the existence of multiple poles (three) in the spectral function of pion, as evidenced by the observations in Figs.~\ref{fig.pola.140}(j) and \ref{fig.pola.180}(i). However, out of these three pion poles, one is unphysical as it gives a negative value of the residue which has been shown by dashed lines in the figure. Further, observing Figs.~\ref{fig.mh}(e)$-$(f),(k) and (l), we notice that at very high values of CCP $\mu_5 \gtrsim 200$ MeV and temperature $T \gtrsim 200$ MeV, both the $\pi$ and $\sigma$ poles posses a negative residue and thus corresponds to unphysical excitations.
Additionally, at higher values of $T$ and $\mu_5$, an abrupt jump-like structure can be seen in the  masses of both the mesonic modes. In order to gain a comprehensive understanding of this phenomenon, in Figs.~\ref{fig.jump}(a)$ - $(c), we have generated spectral function plots for $\pi$ and $\sigma$ as a function of $q^0$  for $\mu_5=150$ MeV at (a) $T = 205$ MeV, (b) $T = 206$ MeV, and (c) $T = 207$ MeV. The solid green and dashed grey vertical lines shows the position of $m_\pi$ and $m_\sigma$ on the $q^0$-axis respectively.
We observe that, both the spectral functions of $\pi$ and $\sigma$ starts from the Unitary cut threshold $q^0=2M$ and they have almost identical shapes, except the fact that the $\pi$ spectral functions has an additional local maximum around the Unitary cut threshold. We find that, the spectral functions posses two maxima separated by a local minima (or dip) which is located at the Landau cut threshold $q^0=2\mu_5$. Similar structure of the spectral function is also present in Figs.~\ref{fig.pola.180}(j)$ - $(l) and Figs.~\ref{fig.pola.220}(h)$ - $(l) i.e. whenever the Landau cut threshold $q^0=2\mu_5$ is larger than the Unitary cut threshold $q^0=2M$. Thus the two maxima separated by a local minima (or dip) in the spectral function of the mesons for $\mu_5>M$ seen here is purely a finite CCP effect. As can be observed from Figs.~\ref{fig.jump}(a)$ - $(c), there is a shift in the global maxima of the spectral function when transitioning from $T=205$ MeV to $T=207$ MeV leading to the observed jump-like structure.

Finally in Fig.~\ref{fig.mott}, we have depicted the variation of the Mott temperature $T_\text{Mott}$ as a function of CCP. We notice that, the Mott temperature lies in the region $180 \lesssim T_\text{Mott} \lesssim 190$ MeV for the entire range of CCP and $T_\text{Mott}$ has mild dependence on $\mu_5$. In particular, $T_\text{Mott}$ possess slightly oscillatory behaviour with $\mu_5$ and shows an overall decreasing trend (with little slope) with the increase in CCP.

%~~~~~~~~~~~~~~~~~~~~~~~~~~~~~~~~~~~~~~~~~~~~~~~~~~~~~~~~~~~~~~~~~~~~~~~~~~~~~~~~~~~~~~~~~~~~~~~
\section{Summary \& Conclusion} \label{sec.summary}
In this work, the neutral meson properties such as mass, polarization functions and spectral functions have been studied in presence of chiral imbalance using two-flavor Nambu--Jona-Lasinio model. To begin with, we have studied the the temperature dependence of the constituent quark mass for different values of CCP. It is found that CCP has the tendency to enhance the chiral condensate in the low temperature region indicating a chiral catalysis. Conversely, at higher temperatures, an opposing effect is observed in which $\mu_5$ weakens the chiral condensate. Consequently, chiral symmetry is restored at a relatively lower temperature compared to the case when $\mu_5=0$, indicating a manifestation of inverse chiral catalysis.

Both real and imaginary part of the polarization function as well as the spectral functions are studied in detail for both scalar and pseudo-scalar channels as a function of $ q^0 $ for different values of CCP  for three different values of temperature capturing different stages of chiral phase transition. A comprehensive analysis of the real part of the scaled polarization function for the $\pi$ meson reveals multiple poles (three) in the propagator of the pseudo-scalar meson. This is observed for specific values of CCP, both in the chiral symmetry-broken phase and in the vicinity of the transition temperature. One of these multiple poles exhibits negative residue and thus corresponds to unphysical mesonic excitation. However, no such effect is observed for the scalar meson. Additionally, the imaginary part of the polarization function for both the $\sigma$ and $\pi$ mesons shows non-trivial Landau cut contributions in addition to the usual Unitary cut contributions. These non-trivial Landau cut contributions are purely a result of finite CCP effects. Moreover, the imaginary part of the polarization function exhibits a highly non-monotonic behavior with respect to $q^0$. At high temperatures, both the real and imaginary parts of the $\sigma$ and $\pi$ mesons are observed to coincide with each other. This merging indicates partial restoration of chiral symmetry. As a consequence of the behavior of the real part of pion polarization function mentioned earlier, at low and intermediate temperatures the spectral function of pions exhibits multiple poles along with the continuum stemming from the two-quark threshold for specific values of CCP. In contrast, the spectral function of sigma mesons shows a Breit-Wigner-like structure that begins from the two-quark threshold, indicating that sigmas are resonant states with finite decay widths. As chiral symmetry is partially restored at higher temperatures, both pions and sigmas appear as resonant states, and their spectral functions become identical. This phenomenon is known as the Mott transition, where the pion spectral function loses its Dirac delta structure and transforms into a Breit-Wigner form similar to that of the sigma meson. The Mott transition temperature shows slightly oscillatory and overall decreasing trend with the increase in CCP. Furthermore, it is observed that an increase in CCP leads to an increase in the width of the spectral function and a decrease in the peak. This implies that both pions and sigmas become more unstable as CCP increases, indicating an enhancement in their decay processes. The variations of pion and sigma masses as functions of temperature as well as CCP is also studied and at higher values of $T$ and $\mu_5$, an abrupt jump-like structure can be seen in the  masses of both the mesonic modes.

%~~~~~~~~~~~~~~~~~~~~~~~~~~~~~~~~~~~~~~~~~~~~~~~~~~~~~~~~~~~~~~~~~~~~~~~~~~~~~
\section*{Acknowledgments}
S.G. is funded by the Department of Higher Education, Government of West Bengal, India. N.C., S.S. and P.R. are funded by the Department of Atomic Energy (DAE), Government of India. 
%~~~~~~~~~~~~~~~~~~~~~~~~~~~~~~~~~~~~~~~~~~~~~~~~~~~~~~~~~~~~~~~~~~~~~~~~~~~~~
 
% \clearpage
%
%\appendix
%\section{Appendix} \label{App.A}

\bibliography{z-Ref}

\end{document}